\documentclass[conference]{IEEEtran}
\IEEEoverridecommandlockouts

\usepackage{cite}
\usepackage{amsmath,amssymb,amsfonts}
\usepackage{graphicx}
\usepackage{textcomp}
\def\BibTeX{{\rm B\kern-.05em{\sc i\kern-.025em b}\kern-.08em
    T\kern-.1667em\lower.7ex\hbox{E}\kern-.125emX}}

\usepackage{xurl}
\usepackage{xspace}
\usepackage[table,xcdraw]{xcolor}

\begin{document}

\newcommand{\OPNAME}{\mbox{DMA-Latte}\xspace}
\newcommand{\OPNAMEBOLD}{\mbox{\textbf{DMA-Latte}}\xspace}
\sloppypar
\newcommand{\authcell}[4]{
  \begin{minipage}[t]{0.32\linewidth}\centering
  {\normalsize #1}\\[3pt]
  {\textit{#2}\\#3\\#4}
  \end{minipage}}

\title{\OPNAME: Expanding the Reach of DMA Offloads\\ to Latency-bound ML Communication\thanks{Our optimized DMA collective prototypes and benchmarking infrastructure are available at \protect\url{https://github.com/AMDResearch/AMD-DMA-Latte}
}}

\author{
\begin{minipage}{\linewidth}\centering
{\normalsize
Suchita Pati\textsuperscript{*}\quad
Shaizeen Aga\textsuperscript{*}\quad
Mahzabeen Islam\textsuperscript{*}\quad
Ryan Quach\textsuperscript{\dag}\quad
Saleel Kudchadker\textsuperscript{*}\quad
Mohamed Ibrahim\textsuperscript{*}}\\[4pt]
{\textit{\textsuperscript{*}Advanced Micro Devices, Inc., USA}\qquad
 \textit{\textsuperscript{\dag}University of California, Riverside, USA}}\\[4pt]
{\{suchita.pati, shaizeen.aga, mahzabeen.islam, saleel.kudchadker, mohamed1.ibrahim\}@amd.com,\quad
ryan.quach001@email.ucr.edu}
\end{minipage}
}

\maketitle

\begin{abstract}

Offloading communication to existing direct memory access (DMA) engines, available on most state-of-the-art commercial GPUs, has emerged as a practical and low-cost solution to efficiently overlap computation and communication in machine learning (ML). However, the reach of DMA offloads has so far been limited to bandwidth-bound scenarios only (10s of MB to GB transfer sizes). In this work, we  break this barrier and extend DMA communication offloads to latency-bound regions (KB to low MB). Specifically, we leverage hitherto untapped features available in the state-of-the-art AMD Instinct\textsuperscript{\texttrademark} MI300X GPUs that render DMA communication offloads competitive even in latency-bound regions. We demonstrate the efficacy of these features both at the operator level (ML communication collectives such as all-gather and all-to-all), and at the end-to-end workload level (LLM inference). At the operator level, our optimizations provide up to 4.5$\times$ speedups (3.2$\times$ geomean in the latency-bound region) over baseline DMA offload, narrowing the performance gap while delivering additional power savings (3-10\%) for ML collectives compared to state-of-the-art GPU core-based communication library, RCCL. At the workload level, we demonstrate acceleration for LLM inference: up to 1.65$\times$ lower latency and up to 1.9$\times$ higher throughput over the state-of-the-art vLLM inference framework. We conclude with a discussion of AMD Instinct GPU runtime innovations that stand to expose these features.

\end{abstract}
\section{Introduction}
\label{sec:intro}

Scaling ML models and training datasets has improved model efficacy and has consequently led to their widespread adoption across a variety of domains. This in turn requires ML training and inference to often be distributed over multiple GPUs, thus necessitating focused optimization of the resulting inter-GPU communication costs. A potential strategy to optimize ML communication overheads is to overlap them with computation by offloading them away from the GPU cores. A low-cost solution for doing this is to offload communication to existing data-movers such as DMA engines. Doing so has been demonstrated to deliver superior computation-communication concurrency by freeing up all GPU cores for computation and to additionally lower interference in the memory sub-system~\cite{agrawal2025optimizingmlconcurrentcomputation,pal2025FiCCO}. Recognizing these benefits, GPU vendors are introducing variants of ML communication collectives that are fully offloaded to GPU DMA engines~\cite{ncclCECollectives}.

\begin{figure}[tb!]
    \centering
    \includegraphics[width=\columnwidth]{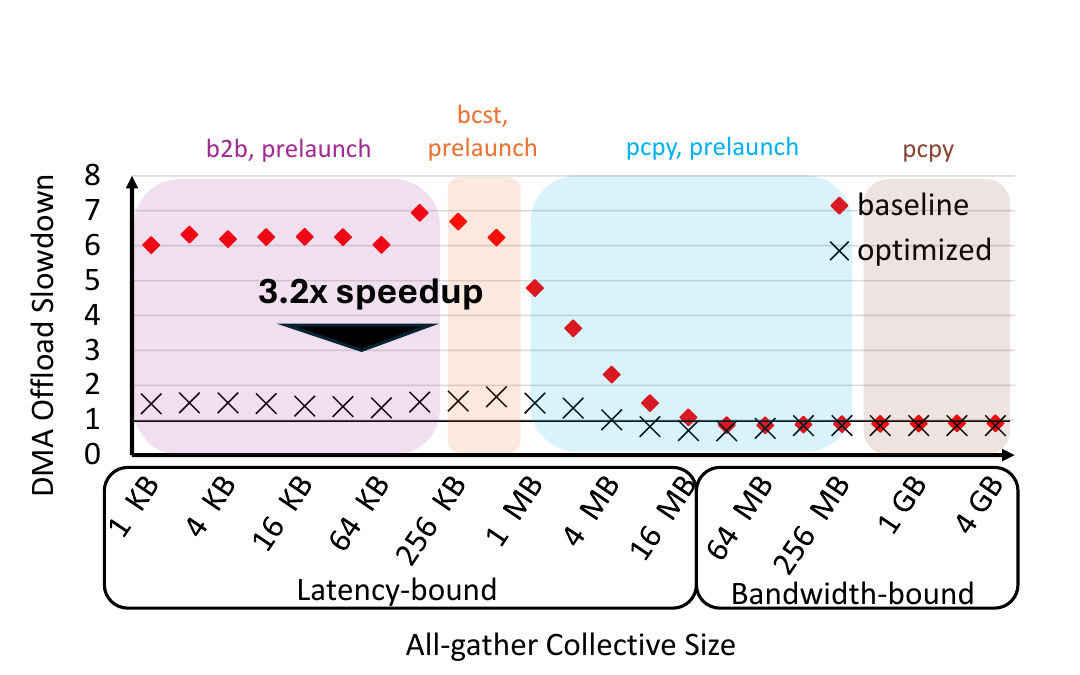}
    \caption{Expanding the reach of DMA offloads to latency-bound ML communication. Proposed \OPNAME harnesses unique DMA features across the size spectrum (b2b: back-to-back copy, bcst: broadcast, pcpy: parallel copy, prelaunch) to provide a 3.2$\times$ geomean speedup across latency-bound sizes.}
    \label{fig:overview}
\end{figure}

Although promising, the reach of DMA offloads has so far been limited to bandwidth-bound scenarios only (10s of MB to GB transfer sizes) as we depict in Figure~\ref{fig:overview}. In the figure, we compare the performance of all-gather ML communication collective when offloaded to DMA engines to that of using state-of-the-art RCCL library~\cite{rccl} which harnesses GPU compute cores or units (CUs). As depicted, for latency-bound scenarios (KB to low MB), DMA collectives considerably lag behind existing collectives libraries (up to 7$\times$ slower). It is this performance gap that has confined DMA offloads to bandwidth-bound regimes, preventing their broader adoption in ML workloads such as context caching~\cite{xie2025strata}, fine-grained computation/communication offloads~\cite{async-tp}, and other latency-sensitive applications.

In this work, we break this barrier and expand the reach of DMA communication offloads to even latency-bound regions (KB to low MB). To do so, we first tease out the reasons for the DMA performance gap by carefully instrumenting the GPU software stack to provide a detailed benchmarking of the latency composition of a single DMA offloaded copy. This detailed benchmarking demonstrates that for latency-bound sizes, DMA command creation, scheduling, and synchronization can be up to 60\% of the total transfer time. 

Next, to directly tackle these overheads, we identify hitherto untapped features available in the state-of-the-art AMD Instinct\textsuperscript{\texttrademark} MI300X GPUs that make DMA communication offloads competitive even for latency-bound regions. First, while existing DMA offload solutions harness only the vanilla \textit{copy} DMA command which copies data from a single source address to a single destination address, we harness two novel DMA commands (\textit{broadcast} and \textit{swap}). As these commands express multiple copy operations with a single command, using them reduces the total number of DMA commands necessary and directly addresses DMA command creation and scheduling overheads. Second, while commercial GPUs have multiple DMA engines and employing multiple engines allows increased parallelism, it also leads to increased scheduling and synchronization costs. We observe that we can harness the ability of DMA engines to efficiently overlap execution of a series of DMA transfers (\textit{back-to-back copy}) to reduce the number of DMA engines employed and thus limiting these overheads. Finally, we also demonstrate the efficacy of \textit{prelaunch}, which allows scheduling of DMA commands off the critical path by triggering them via memory writes. Table~\ref{tab:design_features} summarizes these features and their benefits when designing DMA-based collectives. Figure~\ref{fig:overview} shows how targeted usage of these features stands to close the performance gap of DMA offloads across a spectrum of latency-bound sizes.

\begin{figure}[t!]
    \centering
    \includegraphics[width=\columnwidth]{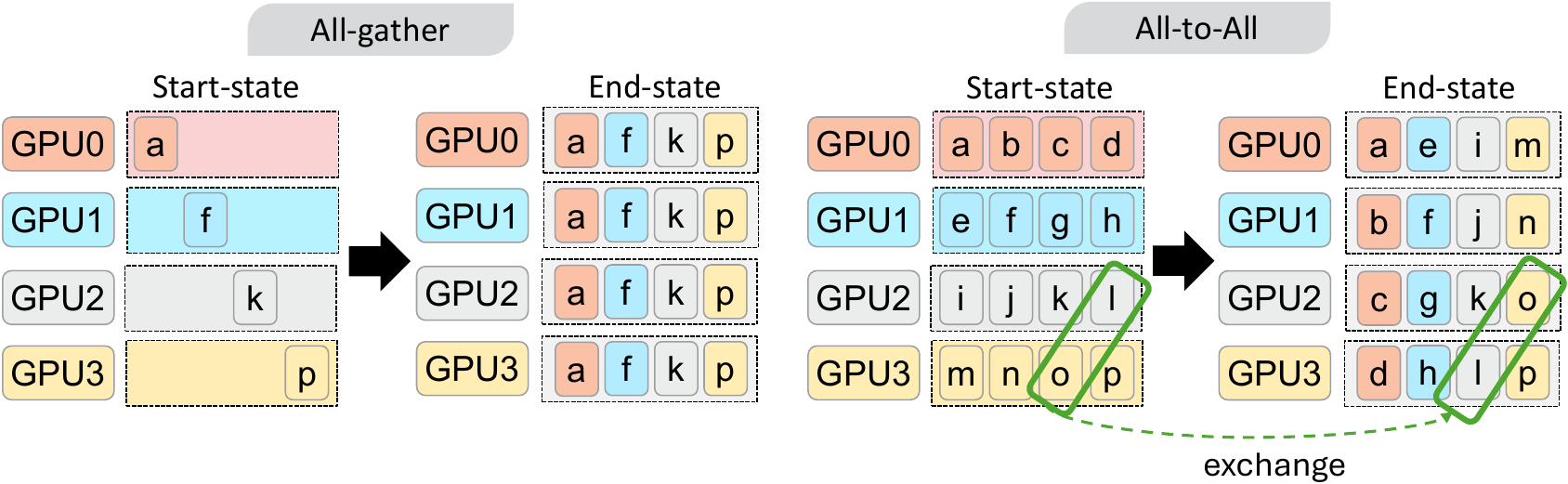}
    \caption{ML collectives: (left) All-gather (right) All-to-All.}
    \label{fig:allgather_alltoall}
\end{figure}

Next, we demonstrate the efficacy of these features on real hardware at both the operator level (ML communication collectives such as all-gather and all-to-all), and the end-to-end workload level (time-to-first-token and tokens/sec). Using our prototypes for all-gather and all-to-all collectives, we demonstrate that we can considerably close the performance gap that DMA collectives face for latency-bound sizes (from 4.5$\times$ and 2.5$\times$
slower to only 30\% slower and 20\% faster all-gather and all-to-all, respectively) 
and deliver additional power savings as well (3-10\%) compared to the state-of-the-art GPU
core-based communication library RCCL.

To demonstrate workload-level benefits, we study key-value (KV) cache save/fetch\footnote{Note that we use the terms KV cache "save" instead of KV cache "offload" in this work to avoid confusion between DMA offloads and KV cache offloads to CPU memory.} to/from CPU memory across inference with a variety of large language models (LLMs). By improving the efficiency of KV cache fetch by DMA engines, we demonstrate up to 1.65$\times$ lower inference latency and 1.9$\times$ higher inference throughput over the state-of-the-art vLLM inference framework.

Finally, while the DMA features we focus on truly stand to expand the reach of DMA offloads and deliver both performance and energy savings, these are not exposed today across the AMD GPU software stack. To address this, we conclude with a discussion of some preliminary work and changes in the AMD Instinct GPU runtime that can expose these features for broader use by the community. 

\begin{figure}[t!]
    \centering
    \includegraphics[width=\columnwidth]{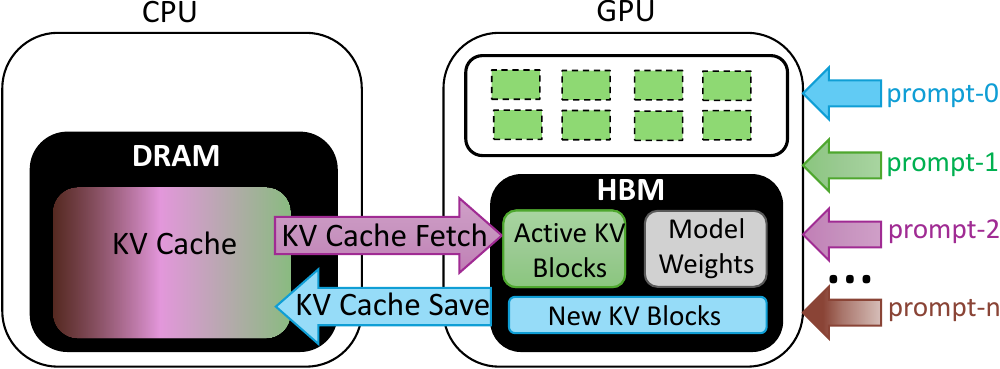}
    \caption{CPU-GPU communication to enable long context caching in ML inference.}
    \label{fig:kv_offload}
\end{figure}

Overall, the key contributions of this work are as follows:
\begin{itemize}
\item We observe in this work that while effective for bandwidth-bound transfers, DMA offloads slow down latency-bound communication by up to 7$\times$, limiting their adoption.

\item 

We carefully instrument the GPU software stack and observe that latency-bound regions manifest high DMA command creation, scheduling, and synchronization overheads, which are up to 60\% of the total transfer time. 
\item Next, we harness hitherto untapped DMA architecture innovations to build both optimized DMA communication collectives (all-gather, all-to-all) and DMA offloads at workload-level (KV cache fetch from CPU memory).
\item Evaluations on real hardware show that 
our optimizations provide up to 4.5$\times$ speedups (3.2$\times$ geomean) over baseline offload for latency-bound collective sizes, narrowing the performance gap and delivering up to 3-10\% additional power savings versus RCCL. Further, our efficient DMA offloads deliver up to 1.65$\times$ lower time-to-first-token latency and 1.9$\times$ higher tokens/sec throughput for inference across a variety of large language models (LLMs). 
\item Finally, we discuss preliminary GPU runtime extensions that can expose the aforementioned DMA architecture innovations for broader use by the community.

\end{itemize}

\section{ML Communication \& Need for DMA Offloads}
\label{sec:bkg_motiv}

\subsection{ML Communication}
\label{sec:bkg_ml_comm}

As model parameters, input datasets and contexts scale, efficiently slicing them across multiple GPU devices while gathering them as needed is crucial for both performance (parallelism) and functionality (memory capacity). Such distributed setups result in communication that manifest either as \textit{collectives} between GPUs or as simple copies between the CPU and GPU.

\begin{figure}[t]
    \centering
    \includegraphics[width=1\columnwidth]{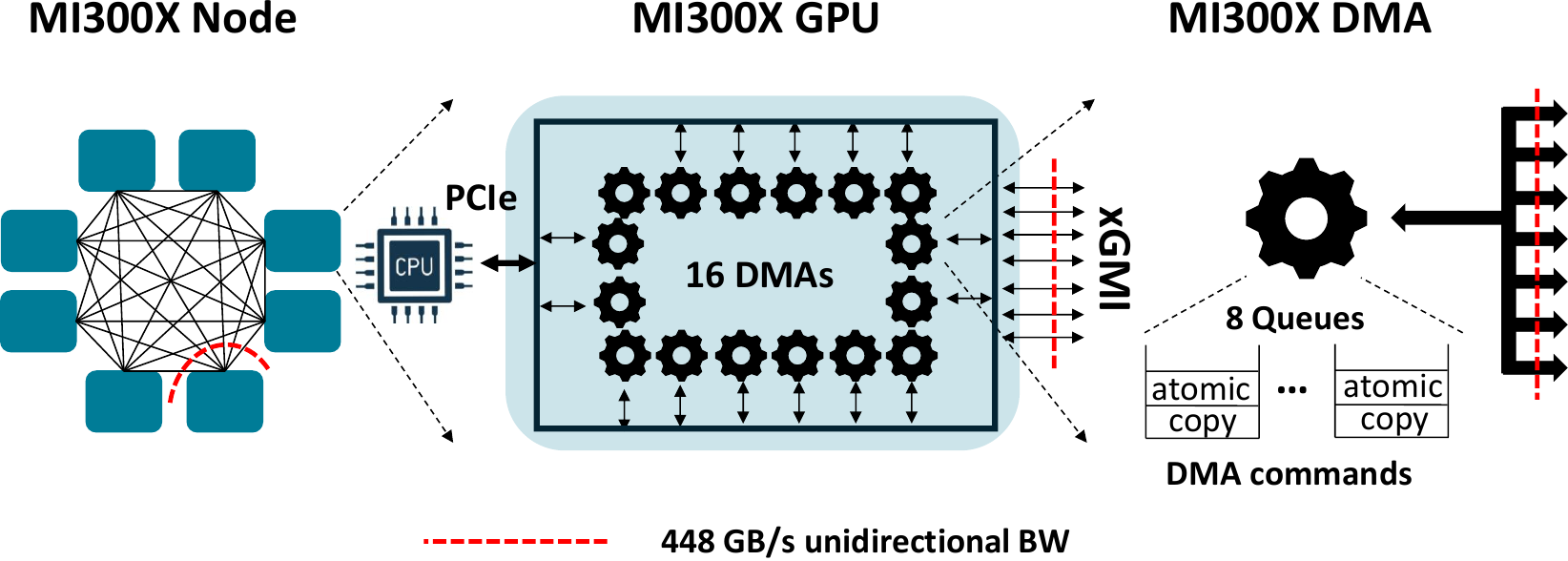}
    \caption{sDMA engines in AMD MI300X Infinity Platform.}
    \label{fig:sdma_background}
    
\end{figure}

\subsubsection{Multi-GPU Collectives}
\label{subsec:bkg_ag_aa}
Techniques such as fully sharded data parallelism (FSDP)~\cite{zhao2023fsdp} and sequence parallelism (SP)~\cite{korthikanti2023reducing} shard weights and/or inputs/activations across multiple GPU devices, and these are aggregated as the model execution progresses. Such aggregation of tensors is achieved by an \textit{all-gather} (AG) collective shown in Figure~\ref{fig:allgather_alltoall} (left), where each GPU starts with a sub-array and receives the remaining sub-arrays from other GPUs. In other words, each GPU sends its respective sub-array to all other GPUs. Another commonly employed collective is \textit{all-to-all} (AA), wherein each GPU starts with a complete array but exchanges sub-arrays with all other GPUs as shown in Figure~\ref{fig:allgather_alltoall} (right). In other words, all participating GPUs collectively perform a transpose operation. Mixture-of-experts (MoE) models in an expert-parallel (EP) setup use AA; each GPU exchanges tokens with other GPUs based on the tokens’ expert preference as determined by the routing layer~\cite{rajbhandari2022deepspeed}. Models also heavily use \textit{reduce-scatter} (RS) collective to reduce gradients from data-parallel model instances and to aggregate partial dot-products in tensor parallelism~\cite{ShoeybiPatwary2019-megatronlm}. RS has similar communication requirements to AA: each GPU sends a distinct sub-array to every other GPU. However, it also requires every GPU to element-wise reduce the received sub-arrays with its local sub-array. 
Since DMAs lack compute support today, not all of RS can be offloaded to DMAs. As such, we focus on optimizing DMA offload of AG and AA collectives.

\subsubsection{GPU-CPU Copies}
\label{subsec:bkg_cpu_gpu}
ML frameworks often also rely on CPU-GPU communication, which has become indispensable for long-context LLM inference. LLM inference has two stages: (i) prefill, which consumes the user prompt plus any existing context (e.g., a document) and generates key–value (KV) tensors, and (ii) decode, which reuses those tensors to produce new tokens without recomputation. The size of these KV caches can grow considerably, in which case frameworks move them to larger but slower memory tiers such as CPU DRAM or SSDs, extending a GPU's effective memory capacity. While this CPU offload of KV cache is similar to those used in training, where optimizer states live on the CPU with gradients transferred from GPU to CPU and updated parameters transferred from CPU to GPU, KV cache offload differs in the number and size of copies that are required when the KV cache is saved to and fetched from CPU memory. This is because most modern LLM frameworks employ a variant of PagedAttention~\cite{kwon2023efficient} which splits the KV cache into small, fixed-size blocks (e.g., 16 tokens in vLLM) stored in non-contiguous memory locations. This reduces memory fragmentation and improves overall memory utilization, allowing more requests to be processed and/or be resident on the system. The trade-off, however, is that a single attention step may need to fetch many independent but dispersed KV blocks, in effect, making these transfers fall in the latency-bound regime. 

\begin{figure}[t]
    \centering
    \includegraphics[width=\columnwidth]{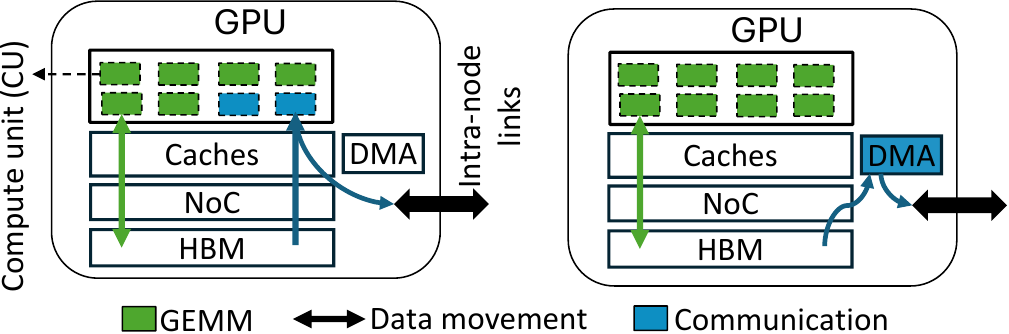}
    \caption{Concurrent compute with communication: (left) using CUs, (right) offloaded to a DMA engine.}
    \label{fig:cu_vs_dma_comm}
\end{figure}

\subsection{DMAs in AMD Instinct\textsuperscript{\texttrademark} MI300X}
\label{subsec:bkg_dma}
State-of-the-art GPU systems have considerable DMA capabilities. As shown in Figure~\ref{fig:sdma_background}, an AMD MI300X Infinity Platform has eight AMD Instinct\textsuperscript{\texttrademark} MI300X
GPUs across which ML collectives are deployed. Each GPU is fully connected with all other GPUs using AMD Infinity Fabric\textsuperscript{\texttrademark} (xGMI) links, each supporting 64 GB/s bandwidth (BW) in each direction, and therefore 128 GB/s bidirectional BW. Thus, each GPU can communicate with all other GPUs with a total BW of 448 GB/s ($7 \times 64$ GB/s) in each direction. 

Each GPU has eight accelerator complex dies (XCDs)~\cite{mi300x-issca}, each of which contains compute units (CUs). The XCDs are vertically stacked over four I/O dies (IODs)~\cite{mi300x-issca,mi300x-vlsi}. The IODs contain the AMD Infinity Cache\textsuperscript{\texttrademark}, the memory interface to the on-package HBM, and 16 system-DMA (sDMA) engines.\footnote{Referred to simply as DMA henceforth.} As shown in Figure~\ref{fig:sdma_background}, each DMA can tackle both CPU-GPU and GPU-GPU communication over PCIe\textsuperscript{\textregistered} Gen 5 and xGMI links, respectively. Note that the next-generation AMD Instinct\textsuperscript{\texttrademark} MI350 and MI355 GPUs have similar DMA capabilities while delivering higher xGMI BW~\cite{rocm_sdma_gfx9}. Although we do not evaluate them in this work, the proposed optimizations apply directly to these newer hardware as well.

Using existing user-facing interfaces, such as the heterogeneous-computing interface for portability (HIP)~\cite{hip_doc} or the heterogeneous system architecture (HSA)~\cite{hsa} runtime API calls, at the user level, a programmer requests a single data transfer to be performed by a DMA engine. More specifically, at HIP level, users invoke \textit{hipMemcpyAsync} API call. If the runtime chooses to use a DMA engine for this copy, the host (CPU) enqueues a \textit{copy} command into the queue of a specific DMA engine. The DMA engine then reads the command, decodes it, and issues the necessary load/store instructions for the copy command. Once the memory accesses complete, the DMA informs the CPU of copy completion using \textit{atomic} commands which increment (or decrement) the value at a provided memory/signal location which the CPU can wait on.

\begin{figure}[tb!]
    \centering
    \includegraphics[width=\columnwidth]{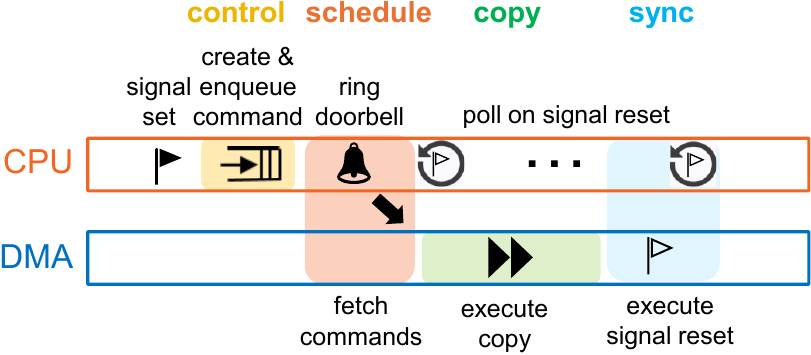}
    \caption{Phases of a single DMA copy offload.}
    \label{fig:dma_phases}
    \vspace{-2.5ex}
\end{figure}

\subsection{Concurrent Compute and Communication}
\label{subsec:bkg_c3}
To reduce communication overheads in distributed ML setups and ensure high scaling efficiency, the communication operations described in Section~\ref{sec:bkg_ml_comm} are often executed concurrently with compute operations~\cite{zhao2023fsdp, rajbhandari2020zero, deepseekai2025deepseekv3technicalreport}. For instance, FSDP overlaps the compute operation of one model layer with the AG of weights from the next layer(s) in a coarse-grained manner~\cite{zhao2023fsdp}. Similarly, Deepseek MoE and vLLM overlap compute and communication across multiple independent instances/inference requests~\cite{deepseekai2025deepseekv3technicalreport,kwon2023efficient}.

In contrast, other mechanisms, such as sequence parallelism (SP) and single-instance MoE, lack independent compute-communication operations, preventing such coarse-grained overlaps. However, several studies have proposed fine-grained overlap techniques~\cite{pati2024t3, pal2025FiCCO}, where tiles or blocks of data generated by a producer kernel are immediately communicated, rather than waiting for the kernel to finish. Overall, overlapping compute and communication is crucial for efficiently scaling ML models.

\subsection{Need for Communication Offloads to DMA}
\label{subsec:bkg_dma_need}

Common GPU implementations of ML collectives from state-of-the-art libraries such as RCCL~\cite{rccl,cowan2023mscclang,nccl} use GPU compute units (CUs) to orchestrate the communication operations. While these implementations are highly optimized for the collectives' isolated execution, they stand to achieve less than ideal performance when overlapped with compute operations~\cite{rashidi2021enabling,agrawal2025optimizingmlconcurrentcomputation}. This is due to interference in both compute and memory subsystem resources as shown in Figure~\ref{fig:cu_vs_dma_comm} (left), which slows down compute and/or collective operations. Offloading communication to other accelerators can help free up compute and cache/memory resources for compute operations and reduce this contention~\cite{rashidi2021enabling,agrawal2025optimizingmlconcurrentcomputation}. Prior work offloads communication to reduce its GPU compute and memory requirements, but proposes a dedicated hardware accelerator to do so~\cite{rashidi2021enabling}. We instead focus on leveraging \textit{existing} DMA engines, available on most state-of-the-art commercial GPUs, to offload these ML collectives and CPU-GPU transfers to alleviate CU and cache contention as shown in Figure~\ref{fig:cu_vs_dma_comm} (right). While dedicated DMA engines are often used for CPU-GPU transfers, recent ML work on CPU offload of KV cache~\cite{xie2025strata} reports high overheads of DMA engines for small-sized transfers and reverts to using kernels/GPU CUs at the cost of the aforementioned contention. In this work, we focus on expanding DMAs' reach to a much wider spectrum of transfer sizes.

\section{DMA Offload: Limits \& Benchmarking}
\label{sec:dma_limits}

\subsection{Bandwidth-bound Coverage}
\label{sec:limits_efficiency}
 In Figure~\ref{fig:overview}, we compare the performance of the all-gather collective when offloaded to DMA engines with that of the state-of-the-art RCCL library~\cite{rccl}, which harnesses GPU compute units (CUs). We also show this for all-to-all collective in Section~\ref{sec:eval}. As depicted, for latency-bound scenarios (KB to low MB), DMA all-gather considerably lags behind existing collective libraries (up to 7$\times$ slower). The same holds for the all-to-all collective (up to 3.6$\times$ slower). This performance gap prevents DMA offloads from being widely used in a variety of ML use cases such as context caching~\cite{xie2025strata}, fine-grained computation/communication offloads~\cite{async-tp}, and more. As such, understanding and closing this performance gap is crucial for broader applicability of DMA offloads' benefits. 

\begin{figure}[tb!]
    \centering
    \includegraphics[width=\columnwidth]{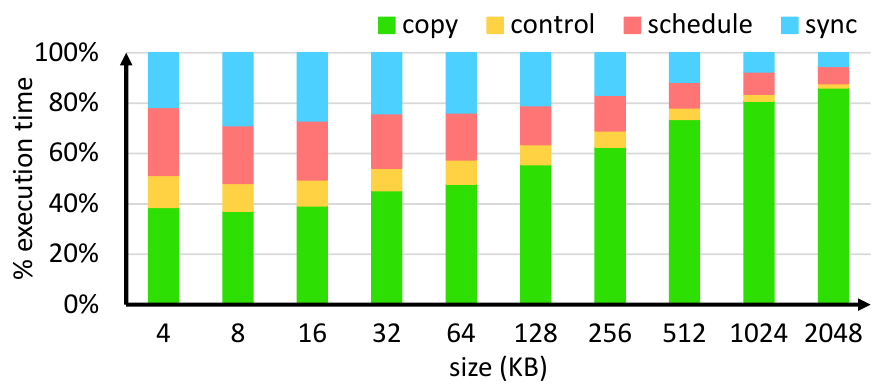}
    \caption{Latency breakdown of a DMA copy.}
    \label{fig:latency_breakdown}
    
\end{figure}

\subsection{Overheads for Latency-bound Sizes}
\label{sec:limits_lb_bech}

In order to expand the reach of DMA offloads, we study the reasons for the lower performance at latency-bound sizes. To do so, as a first step, we study a single DMA copy, which is the smallest unit of work that applications can offload to DMAs with HIP/HSA API calls. We then analyze the distinct phases of a single DMA copy and study where time is expended in order to identify targeted optimization opportunities. 

\subsubsection{Benchmarking Methodology}
\label{sec:limits_meth}
Since much of the DMA functionality resides within the AMD ROCm Runtime (ROCr)~\cite{rocr} (which manages queues, memory, and command submission to the GPU),  we create a micro-benchmark that exercises only the portions of the runtime relevant to a single DMA copy (discussed below in Section~\ref{sec:limits_phases}). 
Like ROCr, we interact with the DMA through the ROC-Thunk interface library (ROCt), which provides user-mode API interfaces to the AMD GPU kernel driver.
This gives us low-level control over commands enqueued in DMA queues, allowing us to insert additional commands such as \textit{timestamp} to tease out time spent in various phases of executing a single DMA copy command. 

\subsubsection{Phases of a DMA Offload}
\label{sec:limits_phases}

Figure~\ref{fig:dma_phases} illustrates 
the phases of a DMA offload for a single copy command. As discussed in Section~\ref{subsec:bkg_dma}, when a user/application issues an asynchronous copy command (hipMemcpyAsync), the CPU creates and enqueues commands to the system-memory-resident request queue for a DMA. We term this first step the \textbf{\textit{control}} phase. The CPU then notifies the DMA of the work by ringing the doorbell, which entails updating the doorbell pointer to point to its write pointer on the queue. A change in the doorbell pointer wakes the DMA engine which then fetches commands from the system memory resident request queue into its internal buffers for processing. We call this CPU to DMA handover as the \textbf{\textit{schedule}} phase. The DMA then executes the copy command, which we call the \textbf{\textit{copy}} phase: it decodes the command, performs address translations, and generates reads/writes from/to the HBM of source/destination devices, respectively. Finally, DMAs use signals (64-bit memory locations) to synchronize with the CPU, notifying it of the completion of the copy. Therefore, a copy is usually followed by a signal update (a memory increment or decrement via an atomic command). We refer to the execution of this command as the \textbf{\textit{sync}} phase. 

\begin{figure}[tb!]
    \centering
    \includegraphics[width=1\columnwidth]{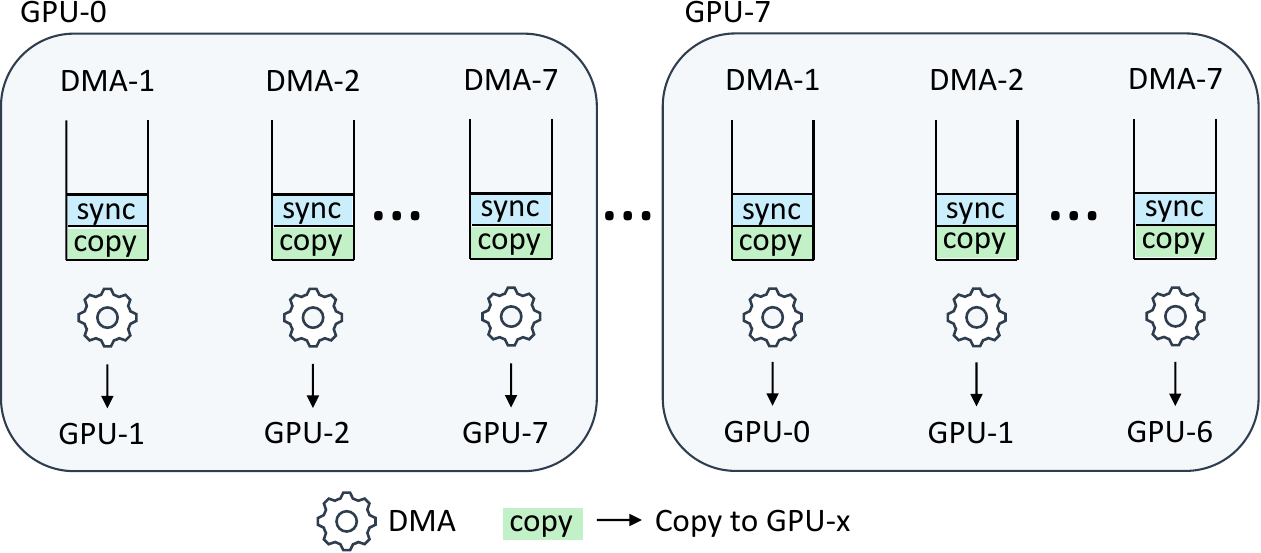}
    \caption{Baseline collective (pcpy) for an eight-GPU system.}
    \label{fig:pcpy}
\end{figure}

\begin{figure}[t!]
    \centering
    \includegraphics[width=1\columnwidth]{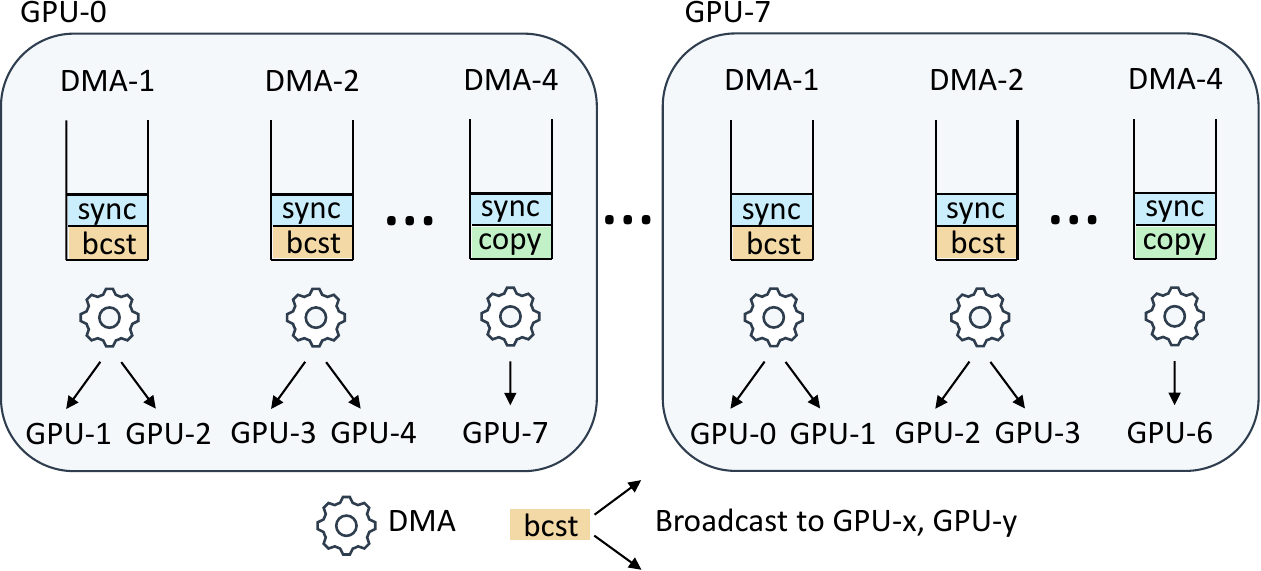}
    \caption{Broadcast-based all-gather for an eight-GPU system.}
    \label{fig:bcst}
\end{figure}

\subsubsection{DMA Copy Breakdown}
\label{sec:limits_breakdown}

Figure~\ref{fig:latency_breakdown} shows the latency breakdown of a single copy operation between two GPUs for a range of sizes (4KB to 2MB). Intuitively, since the cost of the copy phase, that is, the cost of local/remote reads/writes, increases with transfer size while the latencies of other phases remain constant, the proportion of time spent on the copy phase increases with size. Furthermore, across all sizes, the phases can be ordered as copy $>$ schedule $\sim$ sync $>>$ control in terms of their runtime contribution. Notably, the schedule and sync latencies are a considerable portion of DMA copy time. Overall, the non-copy phases --- control, schedule, and sync - account for up to $\sim$60\% of the time at the smallest sizes and are small ($<$20\%) only when copy sizes are $>$1MB. 

ML communication (collectives or otherwise) requires multiple such copy commands. As an example, for all-gather, $n \times (n-1)$ independent copies are required, where $n$ is the number of devices involved in the collective operation. Offloading such ML communication to DMA engines thus requires the host to create several commands, schedule them on queue(s) at one or more DMAs and synchronize with them. These considerably scale the control, schedule, and sync phases and consequently render DMA offloads efficient only for bandwidth-bound sizes as shown by {\tt baseline} in Figure~\ref{fig:overview}. As such, optimizations that target these phases are crucial to expand DMA offload coverage. 

\begin{table*}[t]
\caption{Features harnessed by \OPNAME to expand the reach of DMA offloads to latency-bound sizes.}
\vspace{-2ex}
\label{tab:design_features}
\centering
\renewcommand{\arraystretch}{1.0}
\setlength{\arrayrulewidth}{0.4pt}
\begin{tabular}{|l|c|c|c|c|}
\hline
\rowcolor{gray!25}
 & \textbf{broadcast} & \textbf{swap} & \textbf{back-to-back} & \textbf{prelaunch} \\
\hline
\rowcolor{gray!10}
Lowers \#copy commands?
  & Yes (Lower \#copies)
  & Yes (Lower \#copies)
  & - 
  & - \\
\hline
Lowers \#DMA engines?
  & Yes
  & Yes
  & Yes
  & - \\
\hline
\rowcolor{gray!10}
Lower sync.\ commands?
  & Yes (Fewer \#engines)
  & Yes (Fewer \#engines)
  & Yes (Fewer \#engines)
  & - \\
\hline
Improves link utilization?
  & Yes (1 read / 2 writes)
  & -
  & Yes (Copies overlap)
  & - \\
\hline
\rowcolor{gray!10}
Off-critical path DMA launch?
  & -
  & -
  & -
  & Yes (DMA poll) \\
\hline
Lowers memory traffic?
  & Yes (Source read once)
  & Yes (In place)
  & -
  & - \\
\hline
\rowcolor{gray!10}
Lowers memory capacity?
  & -
  & Yes (In-place)
  & -
  & - \\
\hline
\end{tabular}

\end{table*}

\section{Untapped DMA Potential: Novel DMA Features}
\label{sec:design}

\begin{figure}[t!]
    \centering
    \includegraphics[width=1\columnwidth]{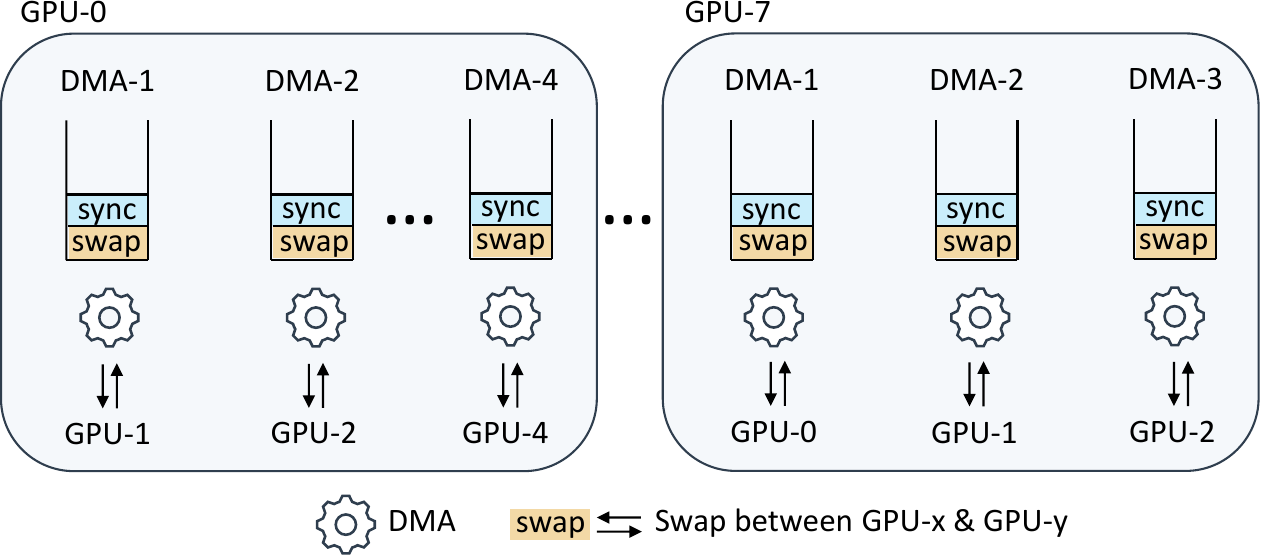}
    \caption{Swap-based all-to-all for an eight-GPU system.}
    \label{fig:swap}
\end{figure}

\begin{figure}[tb!]
    \centering
    \includegraphics[width=0.65\columnwidth]{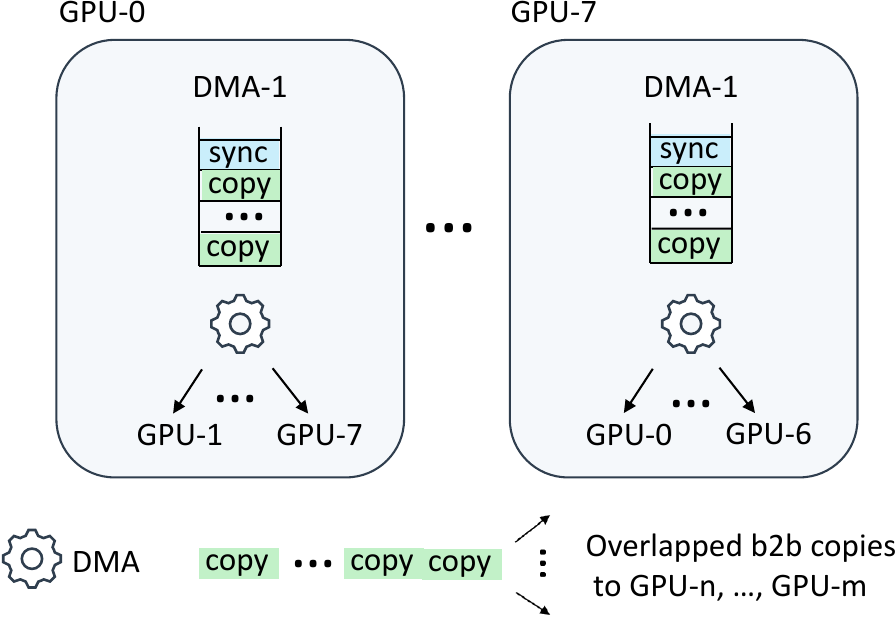}
    \caption{Back-to-back copy-based collective (b2b) for an eight-GPU system.}
    \label{fig:b2b}
    \vspace{-2ex}
\end{figure}

We discuss in this section features available in MI300X GPUs that hold the potential to address the overheads we identified in Section~\ref{sec:dma_limits}. We summarize the features and their benefits in Table~\ref{tab:design_features}. Note that each column in Table~\ref{tab:design_features} lists benefits a given feature can attain in isolation. That said, benefits can be combined across features as we discuss below. As an example, \textit{prelaunch} can be combined with all other listed features.

In order to intuitively explain each feature, we first discuss a baseline ML communication collective design which harnesses DMA offloads and use this straw man to explain each feature's benefits. Note that we use collectives merely as an example and as our workload-level evaluations show (Section~\ref{sec:eval}), these features are broadly applicable (depending on the specific performance bottleneck) for a variety of DMA offload use-cases. 

\subsection{Baseline DMA Offload (\textit{pcpy})}
\label{sec:design_pcpy}

ML collectives such as all-gather and all-to-all require multiple independent GPU-to-GPU transfers (i.e., $n \times (n-1)$, Section~\ref{sec:limits_breakdown}). As each GPU supports multiple DMA engines (Section~\ref{subsec:bkg_dma}), baseline DMA offload solutions spread copies in a single collective over available DMA engines. We term this implementation \textit{pcpy} (parallel copy). In such implementations, minimally, each such DMA engine's queue comprises a \textit{copy} and a \textit{sync} command (assuming \#engines $\geq$ \#independent copies) as shown in Figure~\ref{fig:pcpy}. Both all-gather and all-to-all manifest a similar communication pattern, just with different source buffers (Section~\ref{subsec:bkg_ag_aa}). 

\subsection{Broadcast (\textit{bcst})}
\label{sec:design_bcst}

AMD Instinct\textsuperscript{\texttrademark} MI300X DMA engines support a broadcast command (\textit{bcst}) which allows specifying a single source and two destination memory locations, in contrast to single source/single destination in a vanilla \textit{copy} command that the GPU runtime exposes today. The \textit{bcst} command lowers the number of DMA commands necessary by half for an all-gather collective where a single source buffer is sent to all peers (Figure~\ref{fig:bcst}). Consequently, the \#DMA engines required is halved as well. Since each DMA independently synchronizes with the CPU on completion, this reduction also lowers \#sync commands needed. Furthermore, by reading the source once for both destinations, the \textit{bcst} command allows higher link utilization, especially for latency-bound sizes. Additionally, reusing a read also lowers memory traffic and stands to deliver higher energy efficiency. 
Note that while the \textit{bcst} command is not directly applicable to the all-to-all collective (each transfer uses a unique source buffer), at the application level, mixture-of-experts (MoE) models that employ all-to-all often send a given token to multiple (top-k) experts, for which \textit{bcst} is well suited.
Overall, \textit{bcst} command delivers considerable benefits as listed in Table~\ref{tab:design_features}. 

\subsection{Swap (\textit{swap})}
\label{sec:design_swap}
AMD Instinct\textsuperscript{\texttrademark} MI300X DMA engines also support a swap command (\textit {swap}) which allows swapping the contents of two memory locations. Doing so with the vanilla \textit{copy} command typically requires a temporary buffer and three separate \textit{copy} commands, including to/from the temporary buffer. Such a \textit{swap} command is particularly useful for in-place all-to-all which is effectively a series of swap operations between source buffers of all GPUs as depicted in Figure~\ref{fig:allgather_alltoall} (right). As such, an in-place all-to-all collective can be realized as in Figure~\ref{fig:swap}.

The DMA \textit{swap} command, by virtue of replacing multiple \textit {copy} commands with a single DMA command, has very similar benefits to the \textit {bcst} command as depicted in Table~\ref{tab:design_features}. Additionally, it enables an in-place all-to-all collective, lowering temporary or intermediate buffer requirements. This can have the secondary effect of enabling a larger batch size or reducing the model sharding required, which in turn can lower latency and deliver higher throughput.

\begin{figure}[t!]
    \centering
    \includegraphics[width=\columnwidth]{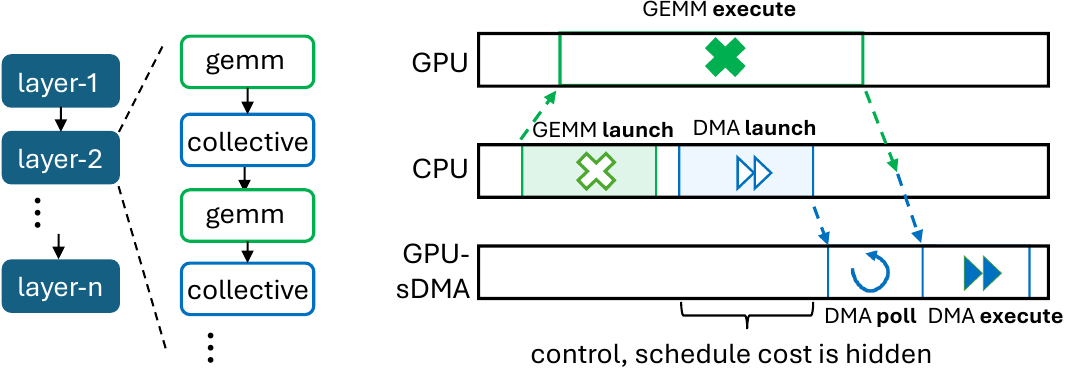}
    \caption{Leveraging deterministic communication patterns in ML to prelaunch collectives.}
    \label{fig:prelaunch}
    
\end{figure}

\subsection{Back-to-back Overlap (\textit{b2b})}
\label{sec:design_b2b}

Another useful feature that AMD Instinct\textsuperscript{\texttrademark} MI300X DMA engines support is overlap of multiple \textit{copy} commands without intervening \textit{sync} operations. For example, when possible, loads of a subsequent copy command can be issued without waiting for stores of a prior copy command to finish, as long as both commands have unique source and destination buffers (no data-dependence hazards). This feature particularly helps small copy sizes by better utilizing the links which otherwise would sit idle due to the time spent in the non-copy phase of small copy commands.

The \textit{b2b} feature is useful in all scenarios where multiple independent copies need scheduling. This includes latency-bound collective sizes as well as latency-bound key-value (KV) cache transfers necessary in context caching~\cite{xie2025strata}. The \textit{b2b} feature has the benefit of using fewer engines, and consequently, fewer commands (lower \textit{sync} commands) as depicted in Figure~\ref{fig:b2b}. 

Note that there exists an interesting tradeoff between harnessing parallelism (more \#engines) and benefiting from \textit{b2b} feature, as the latter requires \textit{copy} commands to be scheduled on a single engine as opposed to former which uses multiple engines for the same task. We leave exploring heuristics for this as future work. 

\subsection{Prelaunch}
\label{sec:design_prelaunch}

While all the above DMA features considerably help lower overheads in each of the phases of the \textit{copy} command, placing command launch on the critical path can degrade DMA offload performance (as is the case with GPU kernel launches as well). In such cases, the \textit{poll} command provisioned by AMD Instinct\textsuperscript{\texttrademark} MI300X DMA engines can be harnessed to pre-schedule or \textit{prelaunch} DMA \textit{copy} commands and take command launch overhead off the critical path. More specifically, pre-scheduled \textit{copy} commands are preceded by the \textit{poll} command which causes the DMA engine to poll a provided memory location until a pre-specified condition is met. With it, DMA commands can be prelaunched without loss of correctness by conditioning their start on completion of their dependencies (prior DMA commands or GPU kernels) via the \textit{poll} command (Figure~\ref{fig:prelaunch}). 

This \textit{poll}-enabled \textit{prelaunch} is particularly relevant for ML communication as communication patterns can often be repetitive. As an example, with fully-sharded data parallelism~\cite{zhao2023fsdp}, all-gather of the next-layer is often scheduled deterministically with computation of the current layer. In such cases, \textit{prelaunch} can take DMA launch overheads off the critical path.

\section{Evaluation}
\label{sec:eval}

\begin{figure}[tb!]
    \centering
    \includegraphics[width=1\columnwidth]{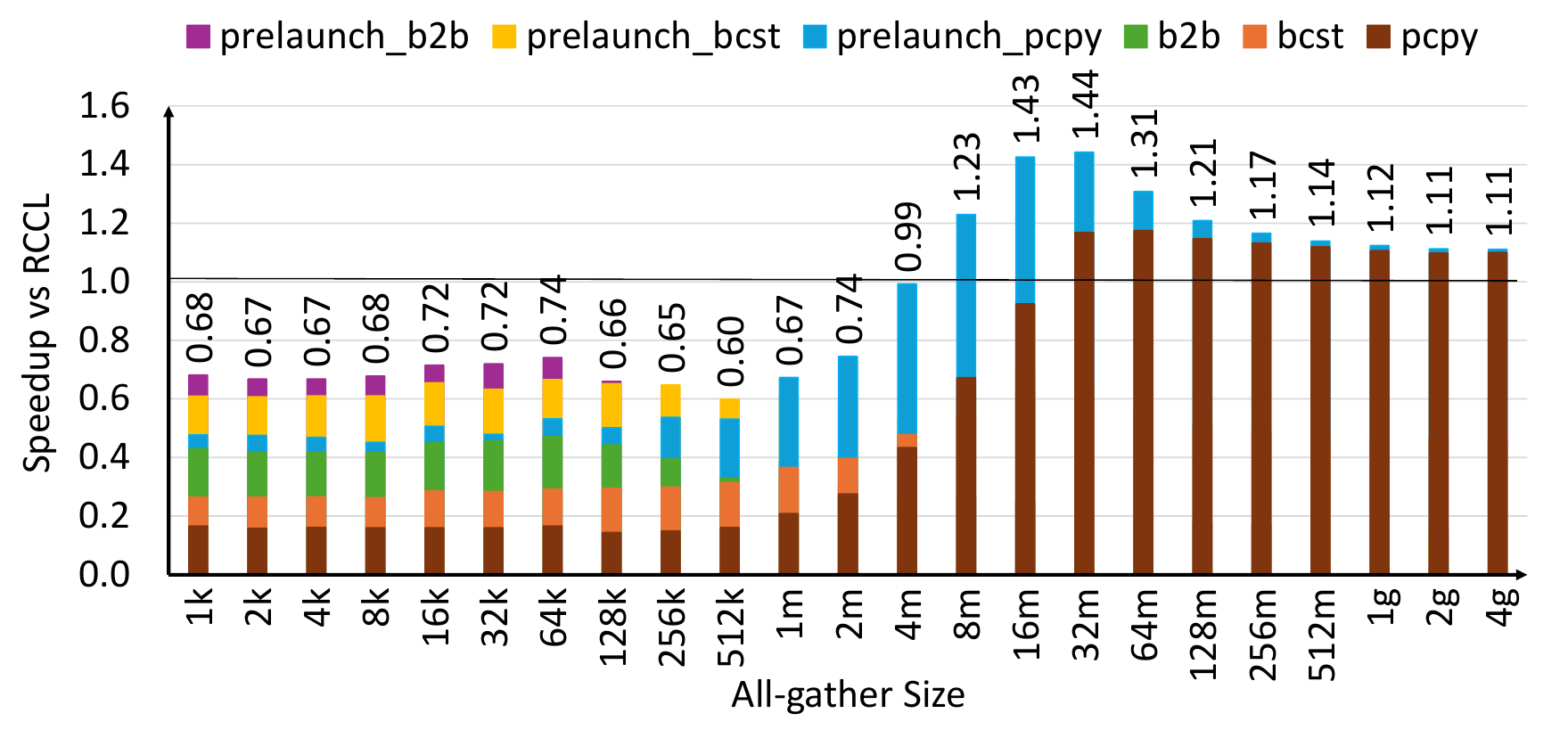}
    
    \caption{Speedup of DMA all-gather collective optimizations versus RCCL.}
    \label{fig:allgather_perf}
    
\end{figure}

\begin{table}[]
\centering
\caption{Performant implementation for all-gather collective.}
\vspace{-2ex}
\label{tab:allgather_map}
{\small
\renewcommand{\arraystretch}{1.0}
\begin{tabular}{|l|c|}
\hline
\rowcolor{gray!25}
\textbf{Size range} & \textbf{DMA Features That Matter} \\ \hline
$1\text{KB} \leq \text{size} < 256\text{KB}$  & b2b, prelaunch  \\ \hline
$256\text{KB} \leq \text{size} < 1\text{MB}$  & bcst, prelaunch \\ \hline
$1\text{MB} \leq \text{size} < 512\text{MB}$  & pcpy, prelaunch \\ \hline
$\geq 512\text{MB}$                            & pcpy            \\ \hline
\end{tabular}
}
\end{table}

We discuss in this section how the features we identify in Section~\ref{sec:design} optimize operator-level performance (ML collectives, Section~\ref{subsec:collective_results}) and how they help at workload-level (KV cache fetch, Section~\ref{subsec:app_results}).

\subsection{System Setup}
\label{subsec:method_setup}
We evaluate DMA-based ML communication on an AMD MI300X Infinity Platform with eight AMD Instinct\textsuperscript{\texttrademark} MI300X accelerators as described in Section~\ref{subsec:bkg_dma} and AMD EPYC\textsuperscript{\texttrademark} 9684X CPUs. The GPUs are fully connected using AMD Infinity Fabric\textsuperscript{\texttrademark} links with bidirectional bandwidth of 128 GB/s. Each GPU is connected to the CPU via PCIe\textsuperscript{\textregistered} Gen 5 links of bidirectional bandwidth 128 GB/s. 

Our software stack uses AMD’s open source ROCm\textsuperscript{\texttrademark}~\cite{rocm_doc}, including the RCCL communication collectives library~\cite{rccl}, the rocBLAS BLAS library for GEMMs~\cite{rocblas}, PyTorch v2.9.0 and a recent (at the time of writing) development version of vLLM (v0.1, commit 5ae65cf0b) for LLM inference evaluations.

\subsection{Closing DMA Collectives Performance Gap for Latency-bound Sizes}
\label{subsec:collective_results}

\subsubsection{Methodology}
We implement our ML collective prototypes using the ROCt library which provides user-level APIs to create DMA packets and interact with DMA engines through the GPU driver. We detail in Section~\ref{sec:software} how the AMD GPU runtime (HIP) has been enhanced to expose some of these DMA features, and our plans to expose the rest. For baseline performance (GPU core or compute-unit/CU  driven collectives), we use AMD's state-of-the-art RCCL~\cite{rccl} library which has been tuned for each message size. We adjust the appropriate environment variables to scale the number of processes, enable performant algorithms (MSCCL~\cite{cowan2023mscclang}, MSCCL++~\cite{shah2025msccl++}) and enable hipGraphs to ensure state-of-the-art baseline performance. 

\begin{figure}[tb!]
    \centering
    \includegraphics[width=1\columnwidth]{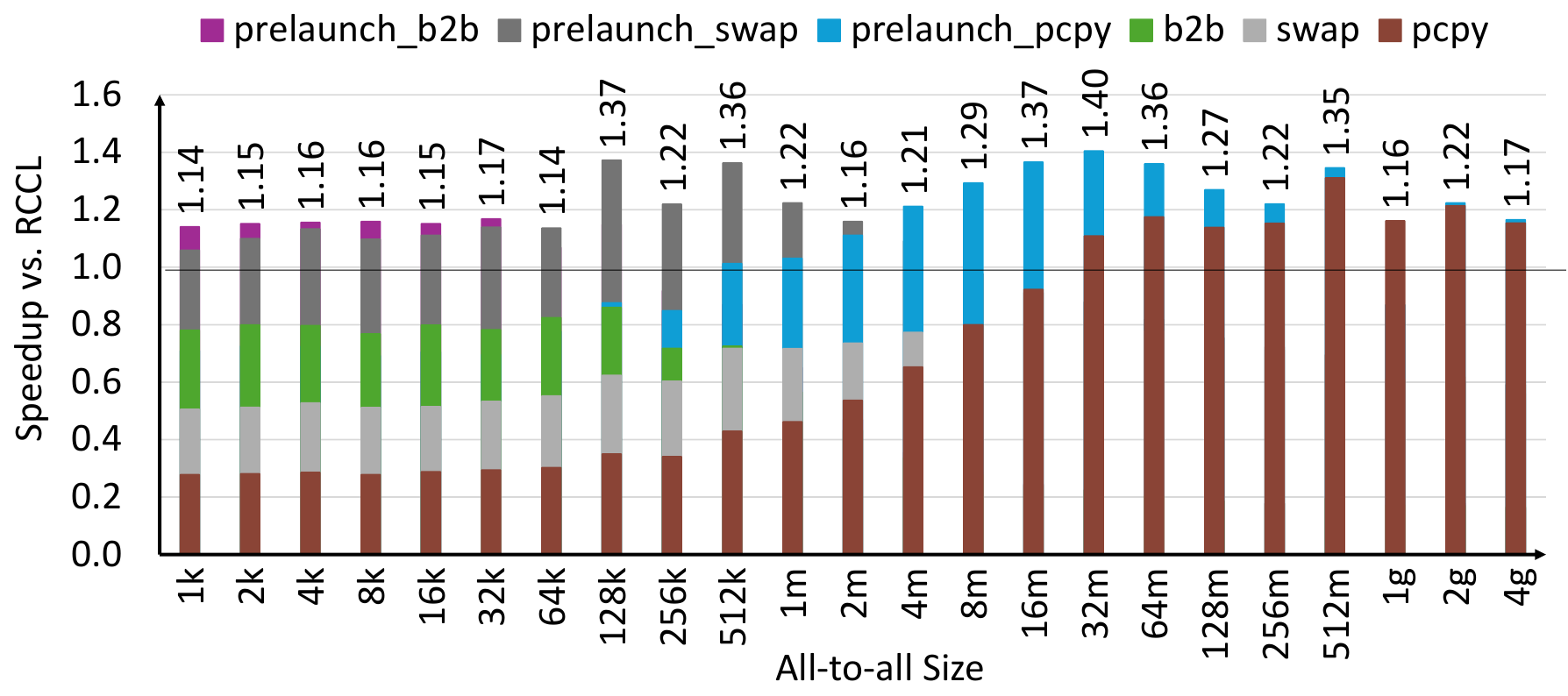}
    
    \caption{Speedup of DMA all-to-all collective optimizations versus RCCL.}
    \label{fig:alltoall_perf}
    
\end{figure}

\begin{table}[]
\centering
\caption{Performant implementation for all-to-all collective.}
\vspace{-2ex}
\label{tab:alltoall_map}
{\small
\renewcommand{\arraystretch}{1.0}
\begin{tabular}{|l|c|}
\hline
\rowcolor{gray!25}
\textbf{Size Range} & \textbf{DMA Features That Matter} \\ \hline
$1\text{KB} \leq \text{size} < 64\text{KB}$  & b2b, prelaunch  \\ \hline
$64\text{KB} \leq \text{size} < 4\text{MB}$  & swap, prelaunch \\ \hline
$4\text{MB} \leq \text{size} < 1\text{GB}$   & pcpy, prelaunch \\ \hline
$\geq 1\text{GB}$                             & pcpy            \\ \hline
\end{tabular}
}
\end{table}

In addition to performance, we also evaluate power behavior of ML collectives. To evaluate power, we use an internal power-logging tool with a sampling interval of one millisecond for averaged power. Samples are captured at different points during the collectives' execution using multiple runs and random delays, following the methodology described in prior work~\cite{singhania2025fingrav}. 

\subsubsection{Collective sizes}
We design and evaluate collectives for a range of sizes that collective libraries support. These include all-gather and all-to-all ranging from 1KB to 4GB. The small (KB) message sizes are important for ML inference scenarios, especially in the LLM decode phases where LLMs operate on/generate a few tokens' worth of activations at a time. These small-sized collectives also manifest prominently with fine-grained compute-communication overlap, requiring a latency-bound collective per step of GEMM execution~\cite{pal2025FiCCO}. The medium and large (MB-GB) collective sizes are common in both training and inference (prefill) and are required for communicating weights and activations depending on the parallelization strategies employed in a given distributed ML setup (e.g., tensor parallelism, expert parallelism). 

\subsubsection{Configurations}
To evaluate our optimized DMA collectives, we compare their performance with the {\tt baseline} or {\tt pcpy} DMA implementations from prior works. But since the goal of DMA collectives is to provide an alternative to CU-based collectives for concurrent compute and communication scenarios (and free up CU and cache resources), we mainly focus on how they perform compared to their best CU-counterpart from RCCL. Thus, in Figures~\ref{fig:allgather_perf} and~\ref{fig:alltoall_perf}, we plot the ratio of CU to DMA collective execution times, or the speedup of DMA collectives over CU-based ones. For both all-gather (AG) and all-to-all (AA), we show \textit{pcpy}, \textit{b2b}, and their \textit{prelaunch} variants. In addition, for AG we show \textit{bcst} and for AA we show \textit{swap} along with their \textit{prelaunch} variants. 

We depict speedups of different variants in a stacked fashion to indicate how we close the performance gap that DMA offloads face when compared to state-of-the-art CU-based collectives. That said, each variant delivers its benefit independent of other variants. For example, for all-gather at 512 KB (Figure~\ref{fig:allgather_perf}), \textit{pcpy} alone achieves only 0.16$\times$ speedup (a 6.25$\times$ slowdown) and \textit{bcst} alone reaches 0.31$\times$ speedup, whereas \textit{prelaunch} combined with \textit{bcst} delivers the best result at 0.6$\times$ speedup (still a slowdown).

\subsubsection{Parallel copy performance (\textit{ pcpy} or \textit{baseline})}
\label{subsubsec:pcpy_perf}
The AG and AA \textit{pcpy} implementation outperforms CU-based collectives by 14\% 
and 18\% geomean for sizes greater than 32MB. This is a result of lower metadata overhead with DMA transfers, which improves network BW efficiency. However, \textit{pcpy} variants are on average 4.5$\times$ 
and 2.5$\times$ slower in the remaining smaller sizes. This is because of the additional non-copy phase times (control, schedule, and sync in Section~\ref{sec:dma_limits}) from engaging several DMA engines that dominate execution at these sizes (eight GPUs, each engaging seven engines for seven peers). While the sync commands (e.g., atomics) execute in parallel, creating and queuing the many sync commands add overheads. Similarly, ringing the doorbell of queues associated with all DMAs also scales such overheads.

\begin{figure}[tb!]
    \centering
    \includegraphics[width=\columnwidth]{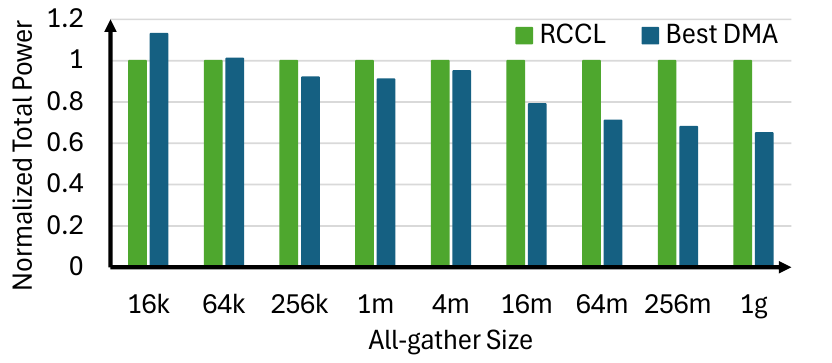}
    \vspace{-4ex}
    \caption{Total power consumed by the best DMA versus CU (RCCL) collectives.}
    \label{fig:power}
    \vspace{-2ex}
\end{figure}

\subsubsection{Broadcast (\textit{bcst}) benefits}
\label{subsubsec:bcst_perf}
As shown in Figure~\ref{fig:allgather_perf}, \textit{bcst} speeds up AG collective by 1.7$\times$ geomean over \textit{pcpy} for sizes up to 4MB, narrowing the gap versus CU-based RCCL from 4.5$\times$ 
to 3$\times$ 
for sizes up to 32MB. This results from using $\sim$half the number of commands and engines, as each broadcast command can transfer to two different destinations (Section~\ref{sec:design_bcst}).
As sizes increase and non-copy phases contribute less to end-to-end time, \textit{bcst}'s benefits decrease. At larger, bandwidth-bound sizes, \textit{bcst} does not provide additional benefits as the parallel engines within each GPU in \textit{pcpy} are able to saturate the total network bandwidth of a single GPU. Overall, while \textit{pcpy} is sufficient for $>$32MB AG collective, \textit{bcst} is required for up to 4MB for improved performance. 

\subsubsection{Swap (\textit{swap}) benefits}
\label{subsubsec:swap_perf}
Similar to \textit{bcst}, \textit{swap} in Figure~\ref{fig:alltoall_perf} speeds up AA collective by 1.7$\times$ geomean over \textit{pcpy} for sizes up to 4MB, and reduces the performance gap versus RCCL from 2.5$\times$ to 1.6$\times$ for sizes up to 32MB. Similar to \textit{bcst}, this results from using half the number of commands and engines as each swap command performs two copies' worth of transfers (Section~\ref{sec:design_swap}). Overall, while \textit{pcpy} is sufficient for $>$32MB AA, \textit{swap} is required for up to 4MB for improved performance. 

\subsubsection{Back-to-back (\textit{b2b}) benefits}
While \textit{bcst} and \textit{swap} help, there is still a considerable 3$\times$ and 1.6$\times$ geomean performance gap between DMA and RCCL collectives for AG and AA, respectively. Our DMA implementation, \textit{b2b}, with all copies from a GPU scheduled back-to-back on a single DMA engine (Section~\ref{sec:design_b2b}) further bridges this gap by speeding up AG collectives by 2.7$\times$ geomean over \textit{pcpy} and by 1.5$\times$ geomean over \textit{bcst} for $<$1MB. Similarly, for AA collectives \textit{b2b} is 2.5$\times$ faster than \textit{pcpy} and 1.4$\times$ faster than \textit{swap} for $<$1MB. Using a single DMA engine reduces non-copy times (fewer sync commands, fewer doorbells) by $\sim$7$\times$ and $\sim$4$\times$ compared to \textit{pcpy} and \textit{bcst/swap}, respectively, with negligible impact on copy times. 

In terms of features, \textit{bcst} and \textit{swap} however remain the superior choice for 1-4MB AG and AA, providing 1.4$\times$ and 1.7$\times$ geomean speedups over \textit{b2b}. Similarly, \textit{pcpy} remains performant at sizes $>$4MB. This is because the increase in total copy time (from interleaving and overlapping seven back-to-back copies) exceeds the benefits from reduced non-copy times in \textit{b2b}. This also demonstrates that across the size spectrum evaluated, each of these features benefits unique size ranges and thus are important. Overall, as shown in Figures~\ref{fig:allgather_perf} and~\ref{fig:alltoall_perf}, including \textit{b2b} reduces the performance gap versus CU-based RCCL from 3$\times$ and 1.6$\times$ to 2.3$\times$ and 1.3$\times$ geomean for sizes up to 32MB for AG and AA.

\subsubsection{Prelaunch benefits}
Finally, we evaluate the impact of prelaunching and hiding the non-copy costs  (Section~\ref{sec:design_prelaunch}). We show its impact on each of the variants; as an example \textit{prelaunch\_b2b} has prelaunch applied on top of the \textit{b2b} feature such that each DMA engine on the GPU has a poll command followed by back-to-back copies and a sync. 

As shown in Figures~\ref{fig:allgather_perf} and~\ref{fig:alltoall_perf}, prelaunching benefits the entire size range. Intuitively, it has the highest impact on implementations which use more commands and DMA engines and have longer non-copy phases; it speeds up \textit{pcpy} by 1.9$\times$, both \textit{bcst} and \textit{swap} by 1.5$\times$, and \textit{b2b} by 1.2$\times$ geomean across the size range. Despite the higher benefit with \textit{pcpy}, prelaunched \textit{b2b}, \textit{bcst} and \textit{swap} still outperform \textit{prelaunch\_pcpy} at sizes $<$ 1MB for AG and $<$ 4MB for AA. The \textit{prelaunch\_pcpy} variant, however, is required for improved performance at larger sizes; it speeds up 1MB-256MB AG by 1.5$\times$ geomean and 4MB-512MB AA by 1.3$\times$ over the best of \textit{pcpy}, \textit{bcst} and \textit{swap}.   

Overall, starting with a 4.5$\times$ and 2.5$\times$ geomean slowdown compared to the state-of-the-art RCCL library for $<$32MB AG and AA, our optimized DMA collectives bring down the slowdown to 30\% geomean for AG and are 20\% faster for AA. They also achieve 20\% speedup over RCCL in the larger, 32MB-1GB, size range. We list the best-performing implementations for different size ranges in Tables~\ref{tab:allgather_map} and~\ref{tab:alltoall_map}. Further DMA optimizations in upcoming AMD GPUs, including packet fusion (using a single packet to express poll, copy, and synchronization instead of three separate packets), could further narrow the remaining performance gap at latency-bound sizes~\cite{dma_packets_2}.

\begin{figure}[tb!]
    \centering \includegraphics[width=1\columnwidth]{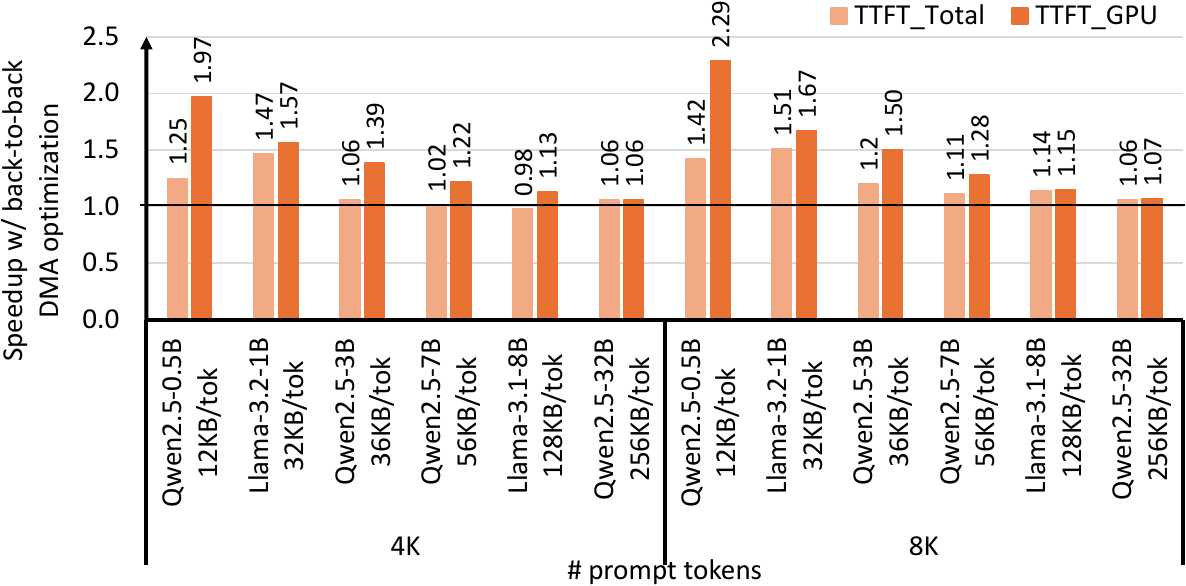}
    
    \caption{TTFT speedups for LLMs with DMA optimizations.}
    \label{fig:ttft_speedup}
    
\end{figure}

\subsubsection{Power Benefits}
\label{subsec:energy}

Offloading communication to DMAs and freeing up GPU compute resources also stands to provide power savings. Figure~\ref{fig:power} shows the total GPU power (including XCD, IOD and HBM  as detailed in Section~\ref{subsec:bkg_dma}) consumed by AG collective and compares the best-performing DMA-offload-based implementation with CU-based ones from RCCL. 

\noindent
\textbf{Bandwidth-bound Sizes:} Figure~\ref{fig:power} shows that DMA collectives consume $\sim$32\% less power than their CU counterparts for larger, $\geq$64MB sizes. This benefit is largely a result of no CU activity (GPU cores) and thus considerably (3.7$\times$) less XCD power with DMA collectives. Since DMA AG is $\sim$20\%  faster than RCCL at these sizes, this translates to energy savings as well.

\noindent
\textbf{Latency-bound Sizes:} DMA offload power benefits decrease at smaller sizes, as RCCL stresses both CUs and memory resources less at these sizes. Nevertheless, our optimized DMA collectives consume less power than the baseline \textit{pcpy} variant (not shown in the figure): \textit{prelaunch\_b2b} lowers power (3–4\% less) by using fewer engines than \textit{prelaunch\_pcpy} in the 16–64KB range, and \textit{prelaunch\_bcst} saves 5–10\% at $>$1MB sizes due to reduced memory traffic (\textit{bcst} reads once and writes to two destinations).   

\subsection{Improved LLM Inference with DMA Offload}
\label{subsec:app_results}

To evaluate end-to-end ML workloads, we use the state-of-the-art vLLM inference framework~\cite{kwon2023efficient}. In particular, we evaluate vLLM's latest KV offload mechanism~\cite{vllm_kv_offload} with our DMA optimizations. Specifically, as fetching KV cache blocks from CPU to GPU memory (Section~\ref{subsec:bkg_cpu_gpu}) involves issuing multiple independent transfers where transfer sizes can be latency-bound, we employ the back-to-back overlap (\textit{b2b}) feature in this case. 

\subsubsection{Implementation}
\label{sec:eval_kv_method}

We use vLLM's KV offload implementation as \textit{baseline} which uses DMA for KV fetch from CPU to GPU~\cite{vllm_kv_offload}. We further modify it to use our \textit{b2b} DMA implementation as detailed below. In addition, we use the kernel-based implementation for KV fetch as in prior works~\cite{vllm_kv_offload} to quantify benefits of reduced contention for GPU cores when using our optimized DMA implementation for KV fetch. 

\noindent
\textbf{Baseline DMA Offloads Implementation}: The baseline vLLM implementation employs \textit{hipMemcpyAsync} calls to transfer KV cache blocks from CPU to GPU memory using DMA. In vLLM, each such KV block contains KV cache pertaining to 16 tokens (vLLM default) but for only one layer. To enable larger, more efficient CPU-GPU copies using DMA, prior work optimized this to store KV cache of all model layers contiguously in memory and we assume this~\cite{vllm_kv_offload}. However, as mentioned in Section~\ref{subsec:bkg_cpu_gpu}, such multi-layer KV cache for different blocks (sets of 16 tokens) are still non-contiguous in memory, requiring the \textit{baseline} configuration to issue multiple DMA-based host-to-device copies (i.e., multiple \textit{hipMemcpyAsync}) to fetch the KV cache for a given request. 

\begin{figure}[tb!]
    \centering
    \includegraphics[width=1\columnwidth]{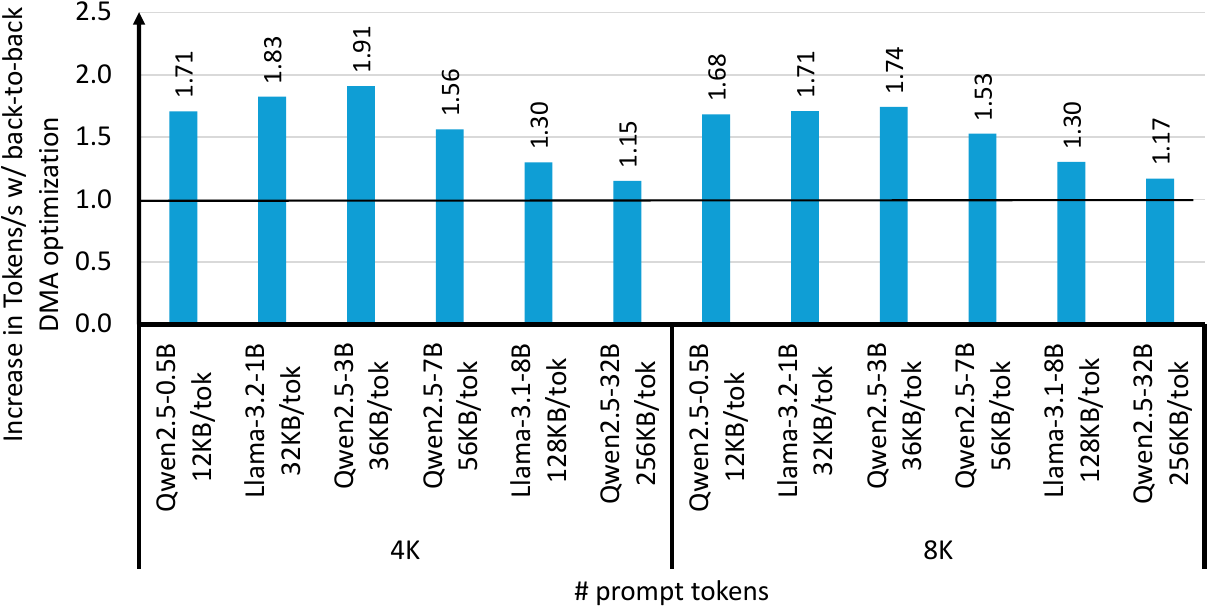}
    
    \caption{Increase in LLM throughput (tokens/s) with optimized DMA offloads.}
    \label{fig:tps_speedup}
    
\end{figure}

\noindent
\textbf{Optimized DMA Offloads Implementation}: In contrast, in our optimized implementation, we first hint to the runtime the presence of independent copies to harness back-to-back overlap (\textit{b2b}) feature. To do so, we harness the \textit{hipMemcpyBatchAsync} API call, which furnishes the runtime with a batch of independent copies (more details in Section~\ref{sec:software}). Further, we modified the HIP and ROCm runtimes~\cite{rocr,clr} to use the \textit{b2b} feature, that is, direct multiple independent copies (about 256 copies in our setup) to a single DMA engine back-to-back with a single synchronization command. In our experiments, we employ a threshold of 4MB based on empirical profiling to pick between \textit{b2b} and harnessing more engines for additional parallelism. 

\noindent
\textbf{Kernel-based Implementation}: We also use the kernel-based KV fetch implementation as in prior works~\cite{vllm_kv_offload}. Instead of launching multiple \textit{hipMemcpyAsync} or \textit{hipMemcpyBatchAsync} API calls to fetch multiple KV blocks using DMA, this implementation launches a single kernel to fetch all dispersed KV blocks using load/store instructions (one kernel workgroup per KV block).

\subsubsection{Methodology}

\label{sec:eval_models_metrics}
We study a spectrum of LLM model sizes along both latency and throughput metrics. 

\noindent
\textbf{LLM Model Setups:} We evaluate popular models deployed today, including Qwen 2.5~\cite{qwen2025qwen25technicalreport,Guo_2025}, Qwen3~\cite{qwen32025} and LLaMa 3.1/3.2~\cite{llama31_8B,llama32_1B}. The model sizes range from 0.5B to 70B (LLaMA-3.1-70B). We use BF16 format weights and all models fit within the 192GB HBM of a single MI300X GPU. We consider prefill lengths ranging from 4K to 128K tokens or up to a model's max supported context length (evaluated Qwen3 models support 32K max context). 

\begin{figure}[tb!]
    \centering
    \includegraphics[width=\columnwidth]{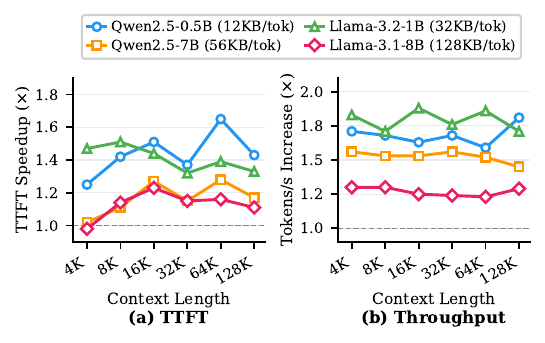}
    \vspace{-5ex}
    \caption{Performance benefits with scaling prompt length.}
    \label{fig:scaling_prompt}
    
\end{figure}

\noindent
\textbf{LLM Inference Metrics:} We evaluate LLM inference performance using two key metrics: (a) time-to-first-token (TTFT) used to measure the response time / latency of a model and (b) tokens per second (TPS) used to measure the throughput of a model. We follow the methodology detailed in prior KV offload evaluations~\cite{vllm_kv_offload}; for TTFT, we fill the CPU memory KV cache with KVs of all tokens and measure the time to generate the first token, given all tokens are cached in CPU memory. We focus on TTFT as KV transfers occur only during prefill. In scenarios where GPU memory accommodates only partial prompt KVs (e.g., contexts exceeding GPU capacity), KV fetches can also occur during decode; our optimizations apply equally, with benefit proportional to the fetch/resident KV ratio. Finally, to evaluate TPS, we use a load of 2000 simultaneous requests with no modification to vLLM's batching policy. We depict performance for scenarios with KV cache hit rate of 100\%, implying that all requests hit in the CPU memory and fetch their KVs from CPU memory. 
This regime is representative of modern multi-turn and agentic serving, where KV-cache reuse across turns is very high: 85-97\% for coding agents~\cite{dynamo_agentic} and above 95\% in recent agentic benchmarks~\cite{agentx}. Nevertheless, we separately discuss the effect of sweeping the KV cache hit rate. That is, a 100\% hit scenario will execute decode only, while hit percentage lower than 100 will execute prefill for the missing KV cache.

\begin{figure}[tb!]
    \centering
    \includegraphics[width=\columnwidth]{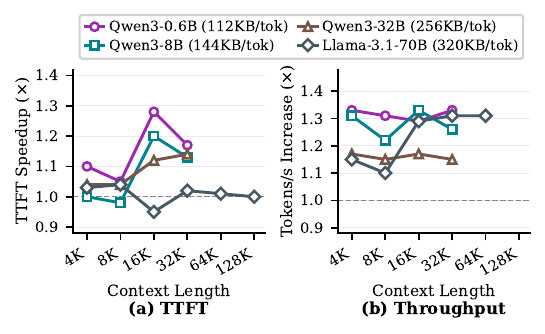}
    \vspace{-5ex}
    \caption{Performance benefits with newer \& larger models.}
    \label{fig:scaling_model}
    
\end{figure}

\subsubsection{KV Fetch Acceleration}
\label{subsubsec:kv_results}
Figures~\ref{fig:ttft_speedup} and~\ref{fig:tps_speedup} show latency and throughput benefits, respectively, of our optimizations on the CPU-to-GPU KV fetch described in Section~\ref{subsec:bkg_cpu_gpu}. 

\noindent
\textbf{Latency (TTFT):} In Figure~\ref{fig:ttft_speedup}, we show that using \textit{b2b} optimization, the total GPU prefill time for model inference/generation (denoted by \textit{TTFT\_GPU}) speeds up by up to 2.29$\times$ over the \textit{baseline} DMA implementation in vLLM. This comes from both reduced overheads from creating fewer synchronization commands when using a single engine --- all independent copies can be scheduled back-to-back with only a single synchronization at the end. In addition, and more importantly, the benefits come from the DMA's ability to overlap the execution of small-sized copies which improves network link bandwidth utilization. Overall, we see benefits are higher for smaller models that have smaller contiguous KV block sizes (calculated as $16~\text{tokens/block} \times \text{KV/token}$, latter is listed per model in Figure~\ref{fig:ttft_speedup}) that suffer from higher overheads and because a larger proportion of GPU execution is spent fetching KV cache (versus model layer execution time). Similarly, the benefits also increase with larger prompt sizes. 

This speedup in GPU execution time along with benefits of reduced API and launch overheads of a single (or few) \textit{hipMemcpyBatchAsync} call result in a total (\textit{TTFT\_total}) speedup of 1.5$\times$. Note that \textit{TTFT\_total} measures the total cost of LLM generation, including all Python, vLLM scheduler and other CPU overheads of launching the GPU execution. 

\noindent
\textbf{Throughput (TPS):} In addition to reduced per-request TTFT, our optimized DMA-offloaded  KV fetch implementation also improves tokens/sec (TPS). Figure~\ref{fig:tps_speedup} shows the TPS of our optimized \textit{b2b} DMA implementation compared to baseline when all (100\%) of prompts/requests hit in the KV cache stored in CPU memory. Our optimization improves TPS by up to 1.9$\times$ when all requests hit in CPU memory. Furthermore, these improvements exceed the TTFT speedups, demonstrating that our DMA optimization not only improves per-response latency over baseline but also improves overlap of the KV fetch with model execution when many such requests are executed concurrently. 

\noindent
\textbf{KV Cache Hit\% Sweep:} We also sweep KV cache hit percentage (50\%, 70\%). In such cases, as GPU kernel execution time increases, benefits of our proposal are expected to drop, which we observe. That said, we also observe that the baseline KV offload implementation does not pick the optimal attention backend for the prefill phase executed when requests miss in CPU memory which further inflates GPU execution time. We leave studying optimized attention backends as future work.

\noindent
\textbf{Scaling Input Length:} Figure~\ref{fig:scaling_prompt} evaluates TTFT and throughput (TPS) as prompts scale to 128K tokens. TTFT benefits generally persist with length — Qwen2.5-0.5B improves from 1.25$\times$ at 4K to 1.65$\times$ at 64K — as longer prompts require proportionally more DMA transfers of the same block size, directly amplifying the benefit of our b2b copy optimizations. Similarly, TPS also improves from 1.71$\times$ to 1.81$\times$ as input scales from 4K to 128K. In some cases (LLaMA) benefits decrease after a certain input length as the attention computation grows to dominate over transfer time. TPS gains remain stable ($\sim$1.7$\times$ for Qwen2.5-0.5B,
  $\sim$1.8$\times$ for Llama-3.2-1B) across all input sizes, indicating improved overlap efficiencies. 

\noindent
\textbf{Evolving Models:} Figure~\ref{fig:scaling_model} evaluates our optimizations on newer and
  larger models. For Llama-3.1-70B, TTFT gains are limited (up to 4\%) as larger hidden dimensions push KV block sizes into the bandwidth-bound regime, leaving less room for b2b copy improvements. Nevertheless, TPS improves by
   up to 1.31$\times$ as DMA efficiency benefits concurrent request execution where compute
  and KV fetch overlap. Note that TPS for 70B at 128K is omitted due to insufficient GPU memory for concurrent requests. Comparing Qwen3 (max 32K context) to its predecessor Qwen2.5, TTFT speedups are smaller (Qwen3-0.6B peaks
  at 1.28$\times$ versus Qwen2.5-0.5B's 1.65$\times$): Qwen3-0.6/8B use 8 KV heads (versus 2 and 4 in Qwen2.5-0.5 and 7B) under GQA,
  increasing the KV block size and reducing the relative benefit of our optimizations. TPS gains also decrease but remain
  substantial at up to 1.33$\times$. Overall, models moving toward leaner attention with smaller KVs will
  benefit more from our optimizations. 

\noindent
\textbf{Scaling Batch-size (B) \& Decode Length:} Increasing batch size has a similar effect on TTFT speedup to increasing prompt length, but with a key distinction: while prompt length scales only KV fetch and attention, batch size scales all layers --- attention, FFN, and embeddings. However, non-attention GEMMs benefit from improved GPU utilization at larger $B$, so their effective cost grows sub-linearly. KV fetch, by contrast, scales linearly with $B$ as block size per request is fixed but we require fetching many more blocks, making our optimizations increasingly impactful at larger batch sizes. For example, LLaMA-3.1-8B TTFT speedup grows from 1.0$\times$ to 1.23$\times$ and Qwen3-32B from 1.04$\times$ to 1.13$\times$ as $B$ scales from 1 to 16. 

For decode, each step reuses the KV fetched during prefill—plus any new KV cached on-GPU from prior decode steps—requiring no additional CPU-to-GPU transfers. Time per output token (TPOT) is therefore unaffected by our optimizations ($\approx$1.0$\times$), and end-to-end speedup decreases as decode length grows: at 16K prefill with 32 decode tokens, Qwen2.5-0.5B achieves 1.13$\times$ end-to-end speedup versus 1.52$\times$ TTFT-only. Nevertheless, as prompt-to-decode ratios increase in agentic workloads, where long context windows are repeatedly queried with short outputs, the prefill phase dominates end-to-end latency, and our optimizations become increasingly impactful~\cite{yuan2026agenticaiworkloadcharacteristics}.

\noindent
\textbf{DMA Versus Kernel KV Fetch:}
Our optimized DMA implementation improves throughput by up to 1.3$\times$ over the kernel-based KV fetch by minimizing resource contention and enabling better compute-communication overlap (Section~\ref{subsec:bkg_dma_need}). That said, the TTFT latency with kernel-based KV fetch is on average 11\% lower than that of DMA-based KV fetch. This is because the former launches a single kernel as opposed to multiple hipMemcpyBatchAsync calls with DMA KV fetch. We leave studying graph launches to reduce launch overheads and optimize latency as future work. 

Overall, our optimized DMA offloads KV fetch implementation improves model response time while also realizing the promise of reduced contention that DMA offloads hold, as shown by the improved throughput.

\section{Runtime Innovations to Expose Novel DMA Features}
\label{sec:software}

\begin{figure}[t!]
    \centering
    \includegraphics[width=\columnwidth]{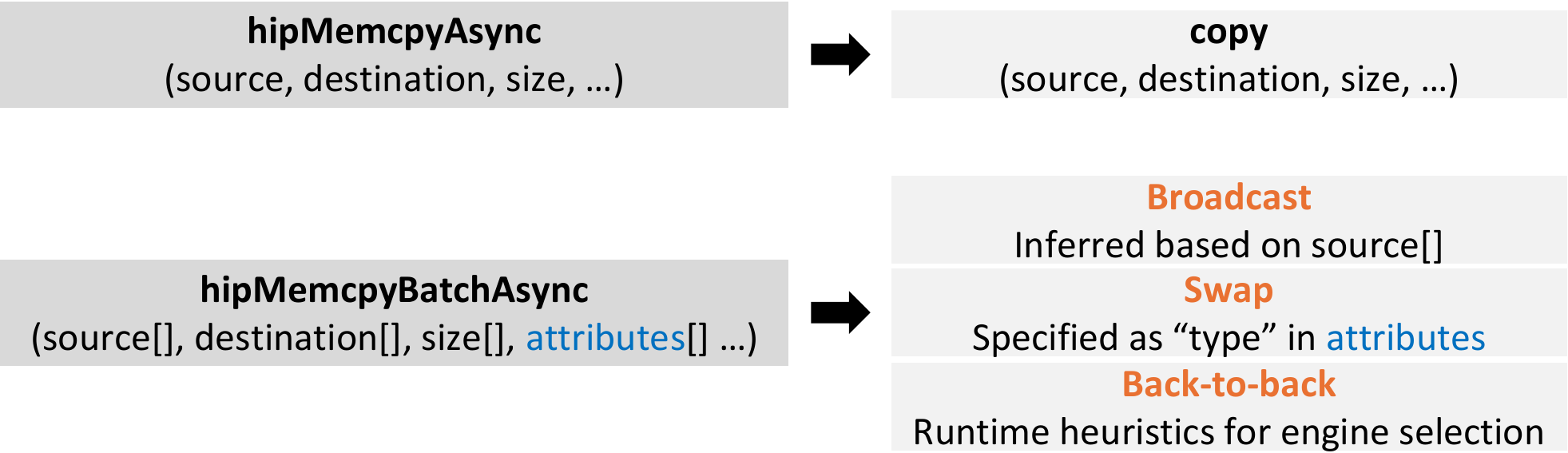}
    \vspace{-3ex}
    \caption{Exposing \OPNAME features in the GPU runtime.}
    \vspace{-4ex}
    \label{fig:sw_features}
\end{figure}

We discuss in this section how we extended the AMD GPU runtime (HIP~\cite{hip_doc}) to expose most of the features we discussed in Section~\ref{sec:design} and possible extensions to expose the rest of the features. While we demonstrate a path to productization using HIP runtime, the implementation can be easily adapted for other DMA runtimes in the presence of the underlying DMA architecture innovations. 

\noindent
\textbf{Copy Batching}: With innovative usage of DMA engines proliferating, GPU vendors have introduced mechanisms to launch multiple asynchronous copies using a single batch API~\cite{hipBatchAsync} in order to contain DMA launch overheads. That is, as depicted in Figure~\ref{fig:sw_features}, while a single \textit{hipMemcpyAsync} can convey a single asynchronous copy to the runtime, \textit{hipMemcpyBatchAsync} conveys a \textit{batch} of asynchronous copies. This mechanism is particularly helpful in providing a rich API which can expose additional DMA features as we discuss below. 

\noindent
\textbf{Broadcast}: With batch API calls like \textit{hipMemcpyBatchAsync}, we extended HIP runtime to inspect an incoming batch of copies and based on source addresses and size of copies, infer if a broadcast operation is desired. In this fashion, transparent to the user, runtime can harness \textit{bcst} DMA command. This also allows the runtime flexibility in picking the command in the regions where it provides benefits (Section~\ref{sec:eval}).

\noindent
\textbf{Back-to-back Overlap}: Existing batch copy API already does not guarantee any specific execution order for copies within the batch. This automatically frees the runtime to opportunistically harness the \textit{back-to-back} overlap optimization. More specifically, for a given GPU system and available IO bandwidth, for transfer sizes falling in latency-bound regions, we extended the HIP runtime to schedule all copies in a batch API call at a single DMA engine (prototyped for KV fetch in Section~\ref{subsec:app_results}), lowering synchronization costs associated with picking multiple DMA engines. 

\noindent
\textbf{Swap}: Unlike broadcast, where the runtime can safely infer if a broadcast is being attempted, doing so with DMA \textit{swap} command is challenging. As such, explicit indication from users is necessary. In such cases, the \textit{attributes} field which provides additional metadata for every copy in batch copy API can provision for a \textit{type} to be associated with every copy. This \textit{type} can then allow users to request a \textit{swap} operation. 

\noindent
\textbf{Prelaunch}: While most of the DMA features we harness in this work can be realized as extensions of \textit{hipMemcpyBatchAsync} API, \textit{prelaunch} requires a slightly different tack. Specifically, as future work, we plan to consider extending graph launch approaches~\cite{hipGraph} to pre-schedule DMA copies by harnessing existing DMA \textit{poll} commands. Further, we also plan to study mechanisms that effectively chain GPU kernel and DMA copies dependencies. Together, we anticipate these to be on par with or exceeding the \textit{prelaunch} gains demonstrated in Section~\ref{sec:eval}.

\section{Related Work}
\label{sec:related}

\noindent
\textbf{Communication Offload: }
Works that improve overlapped compute and communication performance usually offload communication to dedicated accelerators on GPUs, requiring extensive hardware support. Some use custom accelerators, that free up CUs for concurrent compute and reduce memory interference by buffering intermediate data~\cite{rashidi2021enabling}. Similarly, compute-capable switches help offload reduction operations in collectives (such as in allreduce) helping limit traffic over network links and the memory sub-system~\cite{klenk2020network}. Switch offloads, however, still require GPU cores to orchestrate data movement and in-switch commands. We instead offload by leveraging \textit{existing} DMAs, requiring no hardware changes.

A potential challenge with DMA collectives is the inability of existing DMA engines to perform arithmetic operations necessary for reduce-scatter and all-reduce operations. We see this as less of a fundamental challenge than a potential extension to the DMA architecture, should reduction offloads deliver enough returns.

\noindent
\textbf{DMA Offload of ML Collectives:}
Several works offload communication to DMAs. MSCCL++~\cite{shah2025msccl++} initiates DMAs from GPU kernels (through a proxy channel on the CPU) and involves CUs, which we prefer to reserve for compute operations. Furthermore, they implement only the \textit{pcpy} variant for collectives which, along with additional indirection to initiate DMA copies, can hurt KB-MB-sized collective performance. ConCCL~\cite{agrawal2025optimizingmlconcurrentcomputation} shows DMA's potential to accelerate concurrent compute but focuses on large collectives. Similarly, other GPU vendors have also introduced DMA-based collectives that also underperform relative to GPU cores at smaller sizes~\cite{ce_collectives}. In this work, we attempt to expand DMA's reach to the entire spectrum of transfer, and thus collective, sizes by making latency-bound DMA offload efficient. 

\noindent
\textbf{DMA Offload of Other ML Communication:}

Several works leverage DMAs for fine-grained compute-communication overlap~\cite{async-tp,cutlass_dist_gemm,zheng2025tritondistributedprogrammingoverlappingkernels}. Some of these only require a GPU to perform a single peer-to-peer copy at a time rather than a collective which is not performant on all topologies as they can under utilize its network bandwidth~\cite{async-tp,cutlass_dist_gemm}. Others rely on a \textit{pcpy}-like variant that underperforms for KB-MB sizes~\cite{zheng2025tritondistributedprogrammingoverlappingkernels}. Nevertheless, insights from this work can benefit such fine-grained mechanisms involving small-sized transfers/collectives.

Recent works have also evaluated DMA performance for KV cache offload to CPU while highlighting DMA overheads for small-sized/latency-bound transfers required to fetch many non-contiguous KV cache blocks from CPU~\cite{xie2025strata,vllm_kv_offload}. These works either revert to GPU compute cores for performance or make the blocks larger (by storing all layers' KV cache contiguously in memory). We show that the latter can still suffer from overheads that our optimized DMA offload prototypes reduce. Several ML inference techniques require similar state transfers; KV cache transfers between prefill and decode systems in disaggregated setups~\cite{patel2024splitwise}, fetching required MoE weights when offloaded to CPU~\cite{kamahori2024fiddler}, and more. While not evaluated here, our optimizations apply to many such scenarios. 

\noindent
\textbf{DMA Offload of Latency-bound Transfers:}
While some prior works have attempted to reduce DMA offload overheads, they require considerable hardware-software changes. As an example, prior works focus on small-sized DMA transfers by proposing a GPU thread-initiated DMA prototype that requires considerable software and hardware changes~\cite{hwang2023ark}. Other works require hardware modifications to track and trigger compute and DMA communication~\cite{pati2024t3}. This work, in contrast, relies on existing hardware, while demonstrating performance benefits of optimizations on collective operators and end-to-end ML inference on real hardware. It further proposes and prototypes simple API and runtime extensions to expose these optimizations to users/frameworks. 

\noindent
\textbf{DMAs Beyond MI300X GPUs:}

Finally, the DMA commands and features we leverage on MI300X GPUs can be extended to other DMA architectures; other system DMAs~\cite{intel_dsa} already support similar primitives (e.g., broadcast, back-to-back overlap) and could adopt the proposed collective implementations. In the absence of such primitives, similar performance gaps at latency-bound sizes can be expected~\cite{ncclCECollectives}.

\section{Conclusion}
\label{sec:conclusion}

Effectively overlapping and hiding communication is critical for efficient distributed ML training and inference. A low-cost mechanism to do so is harnessing DMA engines available on most state-of-the-art commercial GPUs. We observe in this work that current efforts to harness DMA offloads for ML communication are limited to bandwidth-bound sizes (10s of MB to GB) and cause severe slowdowns in latency-bound regions. To tackle this, we identify hitherto untapped features available in the state-of-the-art AMD Instinct\textsuperscript{\texttrademark} MI300X GPUs that render DMA communication offloads competitive even for latency-bound regions. 
Using prototypes on real hardware, we show these features make DMA offloads compelling for latency-bound sizes. They accelerate ML operators (all-gather and all-to-all collectives) by up to 4.5$\times$ over the baseline DMA offload, and deliver 3-10\% power savings versus RCCL.  At the end-to-end workload level, they reduce time-to-first-token latency by up to 1.65$\times$ and achieve up to 1.9$\times$ higher throughput. We conclude with a discussion of runtime innovations that can expose these features for broader use by the community.

\section*{Acknowledgment}
We would like to thank the anonymous reviewers for their feedback that helped improve this paper. AMD, the AMD Arrow logo, AMD Instinct, AMD ROCm, AMD Infinity Cache, AMD Infinity Fabric, and combinations thereof are trademarks of Advanced Micro Devices, Inc. Other product names used in this publication are for identification purposes only and may be trademarks of their respective companies.

\bibliographystyle{IEEEtran}
\bibliography{text/10-references}

\end{document}